\pdfoutput=1
\documentclass[11pt,letter,twocolappendix]{emulateapj}
\usepackage{amsmath}
\usepackage{natbib}
\usepackage{aas_macros} %needed for ads bib entries to work
\usepackage{verbatim} % multiline comments
\usepackage{multirow} % easy multirow cells

\usepackage{algorithm}
\usepackage{algpseudocode} %float container for algorithm
\usepackage{setspace} %fix tight line spacing
\usepackage{xcolor} 

\newcommand{\lbol}{\mbox{$L_{bol}$}} % bolometric luminosity
\newcommand{\tbol}{\mbox{$T_{bol}$}} % bolometric temperature
\newcommand{\lsun}{\mbox{L$_\odot$}}% Lsun
\begin{document}

\title{Identifying Variability in Deeply Embedded Protostars with ALMA and CARMA}
\author{Logan Francis\altaffilmark{1,2,3}, 
        Doug Johnstone\altaffilmark{1,3}, 
        Michael M. Dunham\altaffilmark{4,5}, 
        Todd R. Hunter\altaffilmark{6}, 
        Steve Mairs\altaffilmark{7}}

\altaffiltext{1}{Department of Physics and Astronomy, University of Victoria, 3800 Finnerty Road, 
Elliot Building, Victoria, BC, V8P 5C2, Canada}
\altaffiltext{2}{loganfrancis3@uvic.ca}
\altaffiltext{3}{NRC Herzberg Astronomy and Astrophysics, 5071 West Saanich Road, Victoria, BC, V9E 
2E7, Canada}
\altaffiltext{4}{Department of Physics, State University of New York at Fredonia, 280 Central 
Avenue, Fredonia, NY, 14063, USA}
\altaffiltext{5}{Harvard-Smithsonian Center for Astrophysics, 60 Garden Street, Cambridge, MA 
02138, USA}
\altaffiltext{6}{NRAO, 520 Edgemont Road, Charlottesville, VA 22903, USA}
\altaffiltext{7}{East Asian Observatory, 660 North A'oh{\=o}k{\=u} Place, University Park, Hilo, 
Hawaii 96720, USA}
\date{Received 2018 September 10; revised 2018 December 7; accepted 2018 December 15; published 
2019 January 29}
 
\begin{abstract}   
    Variability of pre-main-sequence stars observed at optical wavelengths has been attributed to
    fluctuations in the mass accretion rate from the circumstellar disk onto the forming star.
    Detailed models of accretion disks suggest that young deeply embedded protostars should also
    exhibit variations in their accretion rates, and that these changes can be tracked indirectly by
    monitoring the response of the dust envelope at mid-IR to millimeter wavelengths.
    Interferometers such as ALMA offer the resolution and sensitivity to observe small fluctuations
    in brightness at the scale of the disk where episodic accretion may be driven. In this work, we
    present novel methods for comparing interferometric observations and apply them to CARMA and
    ALMA 1.3mm observations of deeply embedded protostars in Serpens taken 9 years apart. We find no
    brightness variation above the limits of our analysis of a factor of $\gtrsim 50\%$, due to the
    limited sensitivity of the CARMA observations and small number of sources common to both epochs.
    We further show that follow up ALMA observations with a similar sample size and sensitivity may
    be able to uncover variability at the level of a few percent, and discuss implications for
    future work.
\end{abstract}

\keywords{accretion, accretion disks - methods: data analysis - stars: formation - stars: 
protostars stars: variables: T Tauri, Herbig Ae/Be - submillimeter: ISM - techniques: 
interferometric }

\section{Introduction}
\label{sec:intro}
 
The earliest stages of star formation begin with the gravitational collapse of a molecular cloud
core to a protostar and circumstellar disk, both surrounded by an extended envelope of gas and dust.
The disk is fed by infall of mass from the envelope, while further growth of the protostar proceeds
by accretion from the disk. The transport of mass through the disk and onto the protostar is
predicted to produce a varying accretion rate  by a wide variety of mechanisms, including
gravitational and/or magneto-rotational instabilities in the disk (e.g., \citealt{armitage2001};
\citealt{vorobyov2005,vorobyov2006,vorobyov2010}; \citealt{machida2011}; \citealt{cha2011};
\citealt{zhu2009a,zhu2009b,zhu2010}; \citealt{bae2014}; \citealt{hartmann2016}), quasi-periodic 
magnetically driven outflows in the envelope \citep{tassis2005}, decay and regrowth of 
magneto-rotational instability turbulence \citep{simon2011}, close interaction in binary systems or 
in dense stellar clusters (\citealt{bonnell1992}; \citealt{pfalzner2008}), and disk/planet 
interactions (\citealt{lodato2004}; \citealt{Nayakshin2012}). 

A variable accretion rate can also resolve the ``luminosity problem"
\citep{kenyon1990,dunham2010,dunham2014}, the discrepancy between the observed luminosities of
protostars, which are spread over several orders of magnitude
\citep{dunham2015,jensenhaugbolle2018} and extend to much lower levels than predicted by models
using a constant accretion rate (e.g. \citealt{shu1987}). If a large portion of a protostar's mass
is accreted in episodes occupying a small fraction ($< 0.01$) of its lifetime, the luminosity
problem vanishes. However, this is not the only plausible solution; a longer protostellar lifetime
with a lower or exponentially decreasing accretion rate %caused by the finite size of the envelope
% or effects of
%outflows
may also resolve or alleviate the luminosity problem \citep{mckeeoffner2011}.

Despite the abundance of possible theoretical origins for variable accretion rates in young stars,
and the likelihood that accretion occurs at a wide range of amplitudes and frequencies (e.g.
\citealt{vorobyov2010}), most observational constraints come from indirect evidence or rarely
observed large amplitude bursts followed over the course of years. The strongest indirect evidence
of variability is found in the clumpy structure of outflows driven by young protostars (e.g.
\cite{plunkett2015}) and their signatures in the envelope chemistry [\citep{taquet2016,rab2017}, see
also reviews in \cite{dunham2014}]. Two examples of types of bursts at optical wavelengths are FUors
and EXors, classes of young T-Tauri stars which are observed to brighten by several magnitudes and
remain bright for decades (FUors) or months to years (EXors) \citep{herbig1977,hartmann1996,
    herbig2008}. This rise in brightness of FUors is interpreted as an increase in the accretion 
    rate
from the disk by factors of $10^2-10^4$ \citep{reipurth1990}, and is believed to occur only a few
times during the formation of a star \citep{audard2014}. More evolved T-Tauri stars are also
inferred to exhibit regular changes in their accretion rate by factors of a few from variations in
emission line strength \citep{costigan2014,venuti2015}.

While changes in the brightness of older T Tauri stars can be monitored in the optical or near-IR,
protostars in the earliest stages of their evolution are too deeply embedded in their nascent
envelopes to be directly observed. At far-IR to mm wavelengths, bulk of the dust in the disk and 
envelope (heated by the protostar) is optically thin to its own emission, and the bolometric
luminosity of the system can be obtained. \cite{johnstone2013} used models of deeply embedded
protostars undergoing sharp increases in their accretion luminosity to show that the envelope heats
up in response to a burst on a timescale of days to months, with the largest and fastest changes in
luminosity occurring at the effective photosphere of the envelope around $\sim 100$ AU. In the
far-IR near the peak of the SED ($\sim 100$ $\mu$m), the observed flux can be used as a direct
measure of the accretion rate as a proxy for the bolometric luminosity. At sub-mm/mm wavelengths,
changes in the flux probe variability in the disk and envelope temperature, resulting in a somewhat
weaker response.

Until recently, only a handful of large amplitude bursts onto deeply embedded (Class 0/I
\footnote{The Class of a young stellar object can be defined by its bolometric temperature
    $T_\text{bol}$ (the temperature of a blackbody with the same mean frequency as the observed
    continuum spectrum.), which is an indirect measure of its evolutionary development. Younger
    protostars have lower $T_\text{bol}$, and ranges of $T_\text{bol}$ divide sources into Class 0
    ($T_{\text{bol}} \le 70$~K), Class I ($70$~K $<T_{\text{bol}} \le 650$~K) and Class II ($650$~K
    $<T_{\text{bol}} \le 2800$~K) \citep{myers1993,chen1995}. Alternatively, the extinction 
    corrected IR spectral index can be used to delineate the classes \citep{greene1994}})
protostars have been detected in the far-IR to mm, and these detections have all been serendipitous.
The protostar HOPS 383 was the first Class 0 protostar found to have undergone an accretion burst,
brightening by a factor of $\sim35$ at 24$\mu$m \citep{safron2015} and a factor of $\sim2$ in the
sub-mm. An outburst at mm wavelengths of a factor of $\sim4$ was found in the massive ($\sim 50-156$
M$_\odot$) and distant ($1.3 \pm 0.09$ kpc) protostellar system NGC 6334-I by comparing 2008
Submillimeter Array (SMA) and 2015 Atacama Large Millimeter/submillimeter Array (ALMA) observations,
corresponding to an increase in luminosity by a factor of $\sim 70$ \citep{hunter2006, hunter2017}.
\cite{liu2017} conducted a 1.3~mm SMA survey of FUors and similar outbursting objects, and very
tentatively detected 30-60\% variability over a period of $\sim$1 year in V2494 Cyg and V2495 Cyg.

The ongoing James Clerk Maxwell Telescope (JCMT) Transient Survey \citep{herczeg2017} is the first
survey designed to monitor for variability in young stellar objects (YSOs) at sub-mm wavelengths.
Eight nearby ($<500$ pc) star forming regions are being monitored at a monthly or better cadence
with the Submillimetre Common-User Bolometer Array 2 (SCUBA-2; \citealt{holland2013}) at 450 and 
850 $\mu$m. As the absolute flux calibration of SCUBA-2 is at the very best
$\sim10\%$ (450 $\mu$m) or $\sim5\%$ (850 $\mu$m) \citep{dempsey2013}, a relative flux calibration
strategy requiring identification and use of stable calibrators in the field is used, and currently
achieves a relative calibration accuracy of $\sim2\%$ at 850 $\mu$m across epochs
\citep{mairs2017a}.

The first half of the 36 month Transient Survey has found that in a sample of 51 protostars brighter
than 350 mJy/beam at 850 $\mu$m (14.6\arcsec beam), 10\% are varying at rates of $\sim\vert
5\vert \%$yr$^{-1}$. Several of the most robust variables are found in the Serpens Main molecular 
clouds, including EC53, SMM10, and SMM1. EC 53 is a Class I protostar already known to be a
variable at 2$\mu$m \citep{hodapp1999,hodapp2012} and which varies at 850 $\mu$m by $\sim$50\% with
an $\sim$18 month period, interpreted as accretion flow mediated by a companion star or planet at
several AU \citep{yoo2017}. The Class 0/I object SMM10 is found to have a fractional increase in
peak brightness of$\sim7\%$yr$^{-1}$ \citep{johnstone2018}. SMM1, a bright intermediate mass Class 0
protostar, is rising in brightness by $\sim5 \%$yr$^{-1}$, \citep{johnstone2018,mairs2017b}, and
0.3\arcsec ALMA observations show it to harbour a high velocity CO jet \citep{hull2016}. HOPS 383,
the serendipitous source in Orion detected by \cite{safron2015}, now appears in decline
\citep{johnstone2018,mairs2017b}.

%  The JCMT beam is 9.8" at 450 $\mu$m
While the Transient Survey monitors a large number of sources over several years, the beam size of
the JCMT at the distances of several hundred pc in the surveyed fields \citep{herczeg2017} includes
much of the outer envelope in the beam, rather than just the effective photosphere surrounding the
disk near $\sim100$~AU where changes in the accretion luminosity are most prominent
\citep{johnstone2013}, thus resulting in dilution of the signal and possible contamination by
heating from the interstellar radiation field. Given that the Transient Survey is still able to find
variations at the level of  $\sim\vert 5\vert \%$yr$^{-1}$, higher resolution observations examining
the disk and inner envelope with similar calibration uncertainty should be more sensitive to
variability. The exquisite resolution and sensitivity provided by interferometric observations with
ALMA is well suited to this task, however, the additional calibration complications caused by
changes in array configuration, spectral setup, and image deconvolution must be taken into account. 
The goal of this work is thus to develop and apply methods for comparing interferometric data in a
variability study of deeply embedded protostars.

For this study, we compare observations of protostars in the Serpens Main molecular cloud observed
at 1.3~mm with The Combined Array for Research in Millimeter-wave Astronomy (CARMA)
\citep{enoch2011} during the 2007 fall, and again 9 years later by ALMA in July 2016 (cycle 3). The
remainder of this paper is organized as follows: In section \ref{sec:obs_and_std_cal}, we describe
our ALMA and the archived CARMA observations of Serpens Main and basic data reduction. In section
\ref{sec:alma_maps} we provide maps of the Serpens Main sample produced from our ALMA data and
summarize the properties of the detected sources. In section \ref{sec:alma_expectations} we compare
our ALMA observations against themselves to determine sensitivity limits of future ALMA observations
for detecting variability. In section \ref{sec:alma_to_carma} we discuss the specialized techniques
required for comparing interferometric observations and present results of a comparison between the
ALMA and CARMA observations. In section \ref{sec:dis_and_conc} we discuss the results and highlight
potential directions for future variability studies.

\section{Observations and Standard Calibrations}
\label{sec:obs_and_std_cal}

Our ALMA observations are designed to measure the 1.3~mm flux of a sample of deeply embedded
protostars for which a previous epoch exists in order to identify any large amplitude ($>50\%$)
variations, and as a baseline for comparisons with future ALMA observations which may uncover much
lower levels of variability. We thus target 12 Class 0 and I sources in the Serpens Main molecular
cloud (table \ref{tab:emb_targets}) previously identified in comparisons of Bolocam and Spitzer 
maps \citep{enoch2009} and mapped at high angular resolution ($\sim1\arcsec$)
with CARMA from 2007-2010 \citep{enoch2011}. One of these sources (SMM1/Ser-emb 6) is a known Class 
0 variable protostar identified by the Transient Survey. 

The Serpens Main star forming region is located $436.0 \pm 9.2$ pc away \citep{ortiz-leon2017} and 
contains
34 Class 0 and I protostars \citep{dunham2015}. The high resolution CARMA maps of Serpens Main
covered the 9 known Class 0 and 3 marginal Class I sources \citep{enoch2009} in order to
constrain the disk and envelope structure of the youngest protostars. Of the 12 sources observed
with CARMA, only 9 were robustly detected in preliminary 110 GHz ($2.7$~mm) and followed up with
230 GHz ($1.3$~mm) observations.

\begin{deluxetable*}{lcccl} %yso_targets
    %\tabletypesize{\tiny}
    %\rotate
    \tablewidth{0pt}
    %\tablenum{}
    \tablecolumns{5}
    \tablecaption{Embedded Protostars observed by ALMA and CARMA}
    \tablehead{\colhead{Source Name} & \colhead{ALMA Pointing Center} & Class\tablenotemark{a} & 
        \colhead{CARMA 230 GHz Map?}  & Other Names} \\
    \startdata
    Ser-emb 1       & 18:29:09.09 +00.31.30.9 & 0 & Y & \\ 
    Ser-emb 2       & 18:29:52.44 +00.36.11.7 & 0 & N & \\ 
    Ser-emb 3       & 18:28:54.84 +00.29.52.5 & 0 & N & \\ 
    Ser-emb 4\tablenotemark{b} (N)   & 18:30:00.30 +01.12.59.4 & 0 & Y & \\ 
    Ser-emb 5       & 18:28:54.90 +00.18.32.4 & 0 & Y & \\ 
    Ser-emb 6       & 18:29:49.79 +01.15.20.4 & 0 & Y & SMM1, FIRS1 \\
    Ser-emb 7       & 18:28:54.04 +00.29.29.7 & 0 & Y &  \\
    Ser-emb 8       & 18:29:48.07 +01.16.43.7 & 0 & Y & S68N \\
    Ser-emb 9       & 18:28:55.92 +00.29.44.7 & 0 & N & \\ 
    Ser-emb 11\tablenotemark{b} (W)  & 18:29:06.61 +00.30.34.0 & I & Y & \\ 
    Ser-emb 15      & 18:29:54.30 +00.36.00.8 & I & Y & \\ 
    Ser-emb 17      & 18:29:06.20 +00.30.43.1 & I & Y &  
    \enddata
    %\tablecomments{}
    \tablenotetext{a}{Division between Class 0 and I determined by \cite{enoch2009}. The analysis 
    of \citep{dunham2015} places all of these protostars in a combined Class "0+I" category.}
    \tablenotetext{b}{Source has multiple components in CARMA observations.}
    \label{tab:emb_targets}
\end{deluxetable*}

\subsection{ALMA Observations and Calibration}

$233$~GHz ($1.3$~mm) Band 6 continuum observations of the Serpens sources in table
\ref{tab:emb_targets} were taken in July 2016 using the ALMA C36-6 configuration to provide
$0.3$\arcsec resolution; further details of the observing setup are listed in table
\ref{tab:alma_obs_setup}. Flux and bandpass calibrators were observed at the beginning of the
schedule, followed by science observations for each target interlaced with (phase) gain 
calibrators. Each science target was observed in two separate scans of equal length (except
Ser-emb 5, which was scheduled with 3 unequal scans) totalling $\sim2$ minutes on source. Automatic
data flagging and flux, gain, and water vapor calibration were applied to the raw visibility data
using the ALMA pipeline in version 4.5.3 of the Common Astronomy Software Applications (CASA)
package\footnote{Available at http://casa.nrao.edu} \citep{mcmullin2007}. In addition to the full
reduction, two subsets of the data were created using only the calibrated science target
visibilities from either the first or second scan (excluding Ser-emb 5) in order to estimate the
detectable lower limits for flux variations for future ALMA observations (see section
\ref{sec:alma_expectations}). Phase-only self-calibration using the CASA \texttt{gaincal} task was 
attempted for every science target in the full data set and each single scan subset. Where 
successful, self-calibration was repeated 2-3 times with successively smaller solution intervals 
ranging from the scan duration to the integration time for each visibility. For the brighter 
targets, self-calibration provided an improvement of up to $30~$\% in SNR,
with a typical increase in peak flux of $\sim5~$\% and reduction in the RMS noise of $\sim5-30~$\%. 

\begin{deluxetable}{ll} %alma_obs_setup
    %\tabletypesize{\tiny}
    %\rotate
    \tablewidth{0pt}
    %\tablenum{}
    \tablecolumns{2}
    \tablecaption{ALMA Observing Setup}
    \tablehead{\colhead{Parameter} & \colhead{Value} }\\
    \startdata   
    Observation date(s)                      &  21 July 2016\\ 
    %more exactly:  21-Jul-2016/03:00:28.0   to   21-Jul-2016/03:37:09.2 (UTC)
    Configuration                            &  C36-6 \\
    Number of Antennas                       &  39 \\
    Project code                             &  2015.1.00310.S \\
    Time per source (minutes)                &  1.75\\
    FWHM primary beam ($\sim 1.13\lambda/D$)(\arcsec) &  22 \\ 
    Proj. baseline range (k$\lambda$)        &  10-808\\
    Resolution (\arcsec)                     &  0.3 \\
    Maximum Recoverable Scale\tablenotemark{b} (\arcsec)     & 3.0 \\
    Sky Frequency (GHz)                      &  233 \\
    Spectral Window Center Freqs. (GHz)      &  224, 226, 240, 242\\
    Channel width (MHz)                      &  15.625\\
    Channels per Spectral Window             &  128 \\
    Total bandwidth (GHz)          &  8\tablenotemark{a}\\
    Flux calibrator                          &  J1751+0939\\
    Bandpass calibrator                      &  J1751+0939\\
    Gain calibrator                          &  J1824+0119
    %    Water Vapor Radiometer calibrators       &  J1819-0258, J1824+0119, J1751+093, 
    %Serpens_emb_2. probably don't need these included but here they are in case someone complains
    \enddata
    \tablenotetext{a}{The bandwidth quoted here is the total before flagging 
    edge channels in each spectral window during calibration. Excluding the flagged edge channels, 
    the bandwidth is 7 GHz.}
    %\tablecomments{More comments here}
    \label{tab:alma_obs_setup}
\end{deluxetable}

\subsection{CARMA Observations and Calibration}

\cite{enoch2011} observed nine of the deeply embedded protostars in Serpens (table
\ref{tab:emb_targets}) at 230 GHz with CARMA, a 23 element interferometer with six $10.4$~m,
nine $6.1$~m, and eight $3.5$~m antennas. The targets were observed using the $10.4$~m and
$6.1$~m antennas from 2007-2010 in CARMA's B, C, D, and E configurations %to sample spatial
% frequencies from 4 to 500 k$\lambda$ \citep{enoch2011}.
to sample spatial scales from $51.6$\arcsec-$0.41$\arcsec. While the maps of the sources 
produced by \cite{enoch2011} combine data from all configurations across three years of 
observations, since we wish to search for variability on month-to year timescales, we instead 
choose to focus on individual $uv$-plane tracks (i.e., nights of observations).

\begin{figure}[htb]
    \centering
    \includegraphics[scale=0.95]{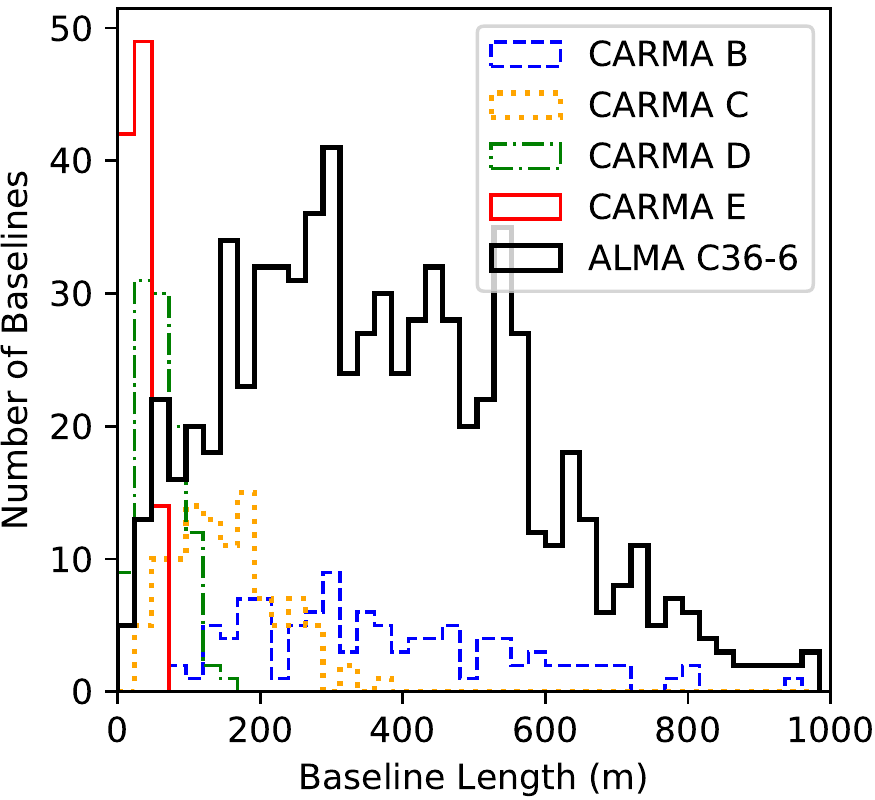}
    \caption{Baseline length distributions for the four CARMA configurations and the ALMA 
        configuration used for our observations. Each bin is 24m wide.}
    %shortest bl for carma c is 30m, shortest for ALMAis 15m    
    \label{fig:bl_comp_kde}
\end{figure}

Each CARMA track samples a range of spatial frequencies in the uv-plane, which are determined by the
(projected) baseline lengths of the array configuration. Ideally, the ALMA and CARMA observations
would have similar $uv$-plane coverage so that the observations would be sensitive to similar
spatial scales of the sky intensity, and we could then produce images and directly compare the
observations. Owing to large differences in the CARMA and ALMA array configurations however, this is
generally not the case. Figure \ref{fig:bl_comp_kde} shows a comparison of the baselines length
distribution in the CARMA B-E configurations and our ALMA configuration. While there is significant
overlap between CARMA B/C and ALMA, the CARMA D and E configurations only have baseline lengths
overlapping with 10-20\% of ALMA. Furthermore, the D and E configurations do not sample the spatial
scales close to the effective photosphere that we wish to compare. While the ALMA data could still
possibly be compared to the CARMA D and E data if most of the ALMA visibilities at large
$uv$-distances were removed (see techniques of section \ref{sec:alma_to_carma}), this would result
in at least a factor of 2-3 drop in the ALMA SNR, and thus we do not attempt to do so.

Although the CARMA B array tracks have $uv$-plane coverage very similar to our ALMA data, the
quality of the data is degraded by worse weather conditions at the wetter CARMA site, and which 
in 
general are poorer for longer CARMA baselines \citep{zauderer2016}. None of the science targets 
in single B-configuration tracks could be detected. We instead concentrate on using the CARMA 
C-configuration data for the variability study (see sections 
\ref{sec:alma_to_carma}.2-\ref{sec:alma_to_carma}.4). 

Several nights of observations in the C-configuration were taken over $\sim2$ weeks in Fall 2007,
the properties of which are summarized in table \ref{tab:carma_c_obs_setup}. Flux and bandpass
calibrators were observed at the beginning or end of each observation followed by interlaced science
and gain calibrator observations. Each track targeted three or four sources for 3-8 hours around
transit (table \ref{tab:carma_tracks_sources}). Integration times varied between sources depending
on the expected flux from single dish observations. The archived raw data were obtained and manually
calibrated using the MIRIAD data reduction package \citep{sault1995}. Once calibration was
accomplished, the data were converted to the CASA measurement set format using the
\texttt{importmiriad} task, and further processing and imaging of the data were carried out in CASA.

%enoch's combined CARMA observations reached a sensitivity of about 1-7 mJy/beam, have 1 \arcsec
%resolution,targets have peak fluxes from 5.6-423 mJy/beam

\begin{deluxetable*}{ll} %carma_c_obs_setup
    %\tabletypesize{\tiny}
    %\rotate
    \tablewidth{0pt}
    %\tablenum{}
    \tablecolumns{2}
    \tablecaption{CARMA C Configuration Observing Setup}
    \tablehead{\colhead{Parameter} & \colhead{Value} }\\
    \startdata   
    Observation date(s)                      &  Oct 24 2007 - Nov 05 2007\\ 
    %more exactly: C1.2: 24-Oct-2007/00:53:59.9 to 24-Oct-2007/03:41:08.8
    %C1.5:  25-Oct-2007/21:19:02.9   to   26-Oct-2007/02:34:23.8 (UTC)
    %C1.8:  26-Oct-2007/23:40:45.4   to   27-Oct-2007/04:27:30.8 (UTC)
    %C2.3   04-Nov-2007/20:01:28.9   to   05-Nov-2007/03:52:38.3 (UTC)
    %Configuration                            &  C \\
    Number of Antennas                       &  6$\times$10.4 m + 8-9$\times$6.1 m \\
    Project code                             &  cx190 \\
    Time per source (minutes)                &  60 - 90\\
    FWHM primary beam\tablenotemark{a} (\arcsec) &  28-47 (37.5) \\
    Proj. baseline range (k$\lambda$)        &  23.1-269.1\\
    Resolution (\arcsec) \tablenotemark{b}                    &  1.5 \\
    Maximum Recoverable Scale (\arcsec)     & 15.5 \\
    Sky Frequency (GHz)                      &  230 \\
    Spectral Window Center Freqs. (GHz)      &  224.0, 224.5, 225.0, 229.5, 230.0, 230.5\\
    Channel width (MHz)                      &  31.25\\
    Channels per Spectral Window             &  15 \\
    Total bandwidth (GHz)                    &  2.8125\\
    Flux calibrator(s)                       &  MWC349, 3C273, Neptune\\
    Bandpass calibrator                      &  J1751+096\\
    Gain calibrator                          &  J1751+096
    \enddata
    %\tablecomments{More comments here}
    \tablenotetext{a}{CARMA's primary beam size varies depending on the combination of 
        10.4 m and 6.1 m antennas used in a baseline. The range from the smallest to largest 
        primary beam sizes is given, and the FWHM of the primary beam that results when data 
        from all baselines is combined is shown in parentheses.}
    \tablenotetext{b}{The resolution listed here is lower than the value calculated from the 
    maximum projected uv-distance due to significant flagging of longer CARMA baselines.}
    \label{tab:carma_c_obs_setup}
\end{deluxetable*}

\begin{deluxetable}{lcccc} %carma_tracks_sources
    %Worth providing Gaussian fits for these here? 
    %\tabletypesize{\tiny}
    %\rotate
    \tablewidth{0pt}
    %\tablenum{}
    \tablecolumns{5}
    \tablecaption{CARMA C Configuration Tracks and Sources Observed}
    \tablehead{\multirow{2}{*}{Field} & \multicolumn{4}{c}{Track Name} \\ %explain shortname?
               & \colhead{C1.2} & \colhead{C1.5} & \colhead{C1.8} & \colhead{C2.3}}\\
    \startdata
    %Track length (s)  & 10028.9 & 18920.9 & 17205.4  & 28269.4
    Track Length (min) & 167 & 315 & 287 & 471 \\ \hline \\
    All fields    & 3 & 3 & 1 & 4 \\ \hline \\
    Ser-emb 1     & 1 & 1 & 1 & - \\
    Ser-emb 2     & - & - & - & - \\
    Ser-emb 3     & - & - & - & - \\
    Ser-emb 4     & 0 & 0 & 0 & - \\
    Ser-emb 5     & 0 & 0 & 0 & - \\
    Ser-emb 6     & 2 & 2 & - & - \\
    Ser-emb 7     & - & - & 0 & 0 \\
    Ser-emb 8     & - & - & - & - \\
    Ser-emb 9     & - & - & - & - \\
    Ser-emb 15    & - & - & - & 1 \\
    Ser-emb 11/17\tablenotemark{a} & - & - & - & 3 
    \enddata
    \tablecomments{0~=~undetected, -~=~unobserved by this track. Note that although Ser-emb 8 was 
        observed at 230 Ghz by \cite{enoch2011}, it was never observed using the C configuration. 
        There is data in the archive for two additional tracks, C1.9 and C1.10, however, they are 
        cut short by degrading weather and no sources can be detected.} 
    \tablenotetext{a}{Observed as a 7-pointing mosaic encompassing both sources.}
    \label{tab:carma_tracks_sources}
\end{deluxetable}

\begin{deluxetable*}{ccccccccc}

    %\tabletypesize{\tiny}
    %\rotate
    \tablewidth{0pt}
    %\tablenum{}
    \tablecolumns{8}
    \tablecaption{ALMA Sources}
    \tablehead{\multirow{2}{*}{ID} & \multirow{2}{*}{Field} & \multirow{2}{*}{Position} & 
    \multirow{2}{*}{Peak 
    Flux} & 
    \colhead{Total Flux} & \colhead{Deconvolved} & \multirow{2}{*}{RMS} & 
    \colhead{Position}\\
    \colhead{} & \colhead{} & \colhead{} & \colhead{} & \colhead{Density} & 
    \colhead{Size\tablenotemark{a}} & \colhead{} & 
    \colhead{Angle}\\
    \colhead{} & \colhead{Ser-emb \#} & \colhead{RA, Dec (ICRS)} & \colhead{(mJy beam$^{-1}$)} & 
    \colhead{(mJy)} 
    & 
    \colhead{(arcsec)} & 
    \colhead{(mJy beam$^{-1}$)} & \colhead{(degrees)} }
    \startdata
    %I've commented out rows where the sources are off center in one field but visibile in another closer to
    %the field center.
1 & 1 & 18:29:09.09 +00:31:30.9 & 92.64 (0.88) & 125.37 (1.90) & 0.24 x 0.14 & 0.20 & 98 &  \\
2 & 2 & 18:29:52.53 +00:36:11.5 & 7.41 (0.30) & 18.95 (1.05) & 0.63 x 0.11 & 0.06 & 165 &  \\
3 & 2 & 18:29:52.54 +00:36:10.3 & 0.97 (0.30) & 0.96 (0.51) & - & 0.06 & - &  \\
4\tablenotemark{b} & 2 & 18:29:52.40 +00:35:52.6 & 8.66 (0.31) & 23.85 (1.12) & 0.59 x 0.26 & 0.06 
& 53 &  \\
5 & 3 & 18:28:54.87 +00:29:52.0 & 9.10 (0.18) & 10.54 (0.35) & 0.15 x 0.09 & 0.07 & 151 &  \\
%6 & 3 & 18:28:55.82 +00:29:44.3 & 3.32 (0.19) & 5.94 (0.51) & 0.30 x 0.26 & 0.07 & 107 &  \\
%7 & 3 & 18:28:55.77 +00:29:44.1 & 2.94 (0.19) & 5.17 (0.49) & 0.36 x 0.20 & 0.07 & 75 &  \\
6 & 4 (N) & 18:29:59.94 +01:13:11.3 & 2.80 (0.11) & 3.66 (0.24) & 0.22 x 0.12 & 0.07 & 51 &  \\
7 & 4 (N) & 18:30:00.67 +01:13:00.1 & 3.16 (0.11) & 3.50 (0.21) & 0.13 x 0.07 & 0.07 & 68 &  \\
8 & 4 (N) & 18:30:00.73 +01:12:56.2 & 3.14 (0.11) & 3.43 (0.20) & 0.12 x 0.06 & 0.07 & 131 &  \\
9 & 5 & 18:28:54.91 +00:18:32.3 & 7.85 (0.09) & 10.08 (0.19) & 0.19 x 0.12 & 0.07 & 162 &  \\
10\tablenotemark{c} & 6 & 18:29:49.80 +01:15:20.3 & 342.46 (5.35) & 985.14 (20.05) & 0.45 x 0.40 & 
0.67 & 165 &  \\
11\tablenotemark{c} & 6 & 18:29:49.66 +01:15:21.1 & 29.33 (4.98) & 119.72 (24.85) & 0.68 x 0.45 & 
0.67 & 88 &  \\
12 & 7 & 18:28:54.06 +00:29:29.3 & 16.75 (1.02) & 22.12 (2.16) & 0.20 x 0.16 & 0.08 & 77 &  \\
13 & 8 & 18:29:48.72 +01:16:55.5 & 15.19 (1.48) & 37.28 (4.92) & 0.48 x 0.29 & 0.13 & 69 &  \\
14 & 8 & 18:29:48.09 +01:16:43.3 & 28.81 (1.51) & 53.18 (4.04) & 0.33 x 0.24 & 0.13 & 35 &  \\
%17 & 9 & 18:28:54.87 +00:29:52.0 & 8.79 (0.20) & 10.60 (0.39) & 0.20 x 0.05 & 0.07 & 141 &  \\
15 & 9 & 18:28:55.82 +00:29:44.3 & 3.34 (0.20) & 5.07 (0.47) & 0.29 x 0.17 & 0.07 & 99 &  \\
16 & 9 & 18:28:55.77 +00:29:44.1 & 3.14 (0.21) & 5.35 (0.52) & 0.29 x 0.23 & 0.07 & 48 &  \\
17 & 11 (W) & 18:29:06.62 +00:30:33.9 & 30.77 (0.54) & 57.20 (1.45) & 0.30 x 0.28 & 0.14 & 87 &  \\
18 & 11 (W) & 18:29:06.77 +00:30:34.1 & 16.35 (0.52) & 20.89 (1.08) & 0.19 x 0.13 & 0.14 & 163 &  \\
%22 & 11 (W) & 18:29:06.20 +00:30:43.0 & 40.68 (0.54) & 94.06 (1.71) & 0.38 x 0.33 & 0.14 & 138 &  
%\\
%23 & 11 (W) & 18:29:05.61 +00:30:34.8 & 7.21 (0.50) & 7.30 (0.88) & - & 0.14 & - &  \\
19 & 11 (W) & 18:29:07.09 +00:30:43.0 & 3.03 (0.47) & 2.49 (0.72) & - & 0.14 & - &  \\
20 & 15 & 18:29:54.30 +00:36:00.7 & 34.58 (0.53) & 61.48 (1.40) & 0.41 x 0.15 & 0.07 & 117 &  \\
%26 & 17 & 18:29:06.62 +00:30:33.9 & 29.14 (0.63) & 54.95 (1.70) & 0.31 x 0.28 & 0.12 & 118 &  \\
%27 & 17 & 18:29:06.77 +00:30:34.1 & 15.63 (0.60) & 19.56 (1.22) & 0.20 x 0.09 & 0.12 & 151 &  \\
21 & 17 & 18:29:06.20 +00:30:43.0 & 41.48 (0.62) & 97.87 (2.00) & 0.38 x 0.35 & 0.12 & 138 &  \\
22 & 17 & 18:29:05.61 +00:30:34.8 & 7.17 (0.58) & 7.76 (1.07) & 0.11 x 0.06 & 0.12 & 165 &  
%30 & 17 & 18:29:07.09 +00:30:43.0 & 2.79 (0.53) & 2.23 (0.81) & - & 0.12 & - &
    \enddata
    %\tablecomments
    \tablenotetext{a}{A ``-'' in the deconvolved size column indicates the source  is un-resolved.}
    \tablenotetext{b}{This source is located $\sim19$\arcsec~from the pointing 
    center, where the response of the primary beam in CASA is 0.188. The primary beam correction is 
    not well modelled this far from the pointing center, and may be uncertain by a factor of $\sim 
    2$. See NAASC memo 117 for a detailed discussion: 
    \url{http://library.nrao.edu/public/memos/naasc/NAASC\_117.pdf}.}
    \tablenotetext{c}{Associated with SMM1}
    \label{tab:alma_sources}
\end{deluxetable*}

\section{Reduced ALMA Maps and Source Identification}
\label{sec:alma_maps}

\begin{figure*}[htb] %alma_serpens_maps_1
    \centering
    \includegraphics[scale=1.0]{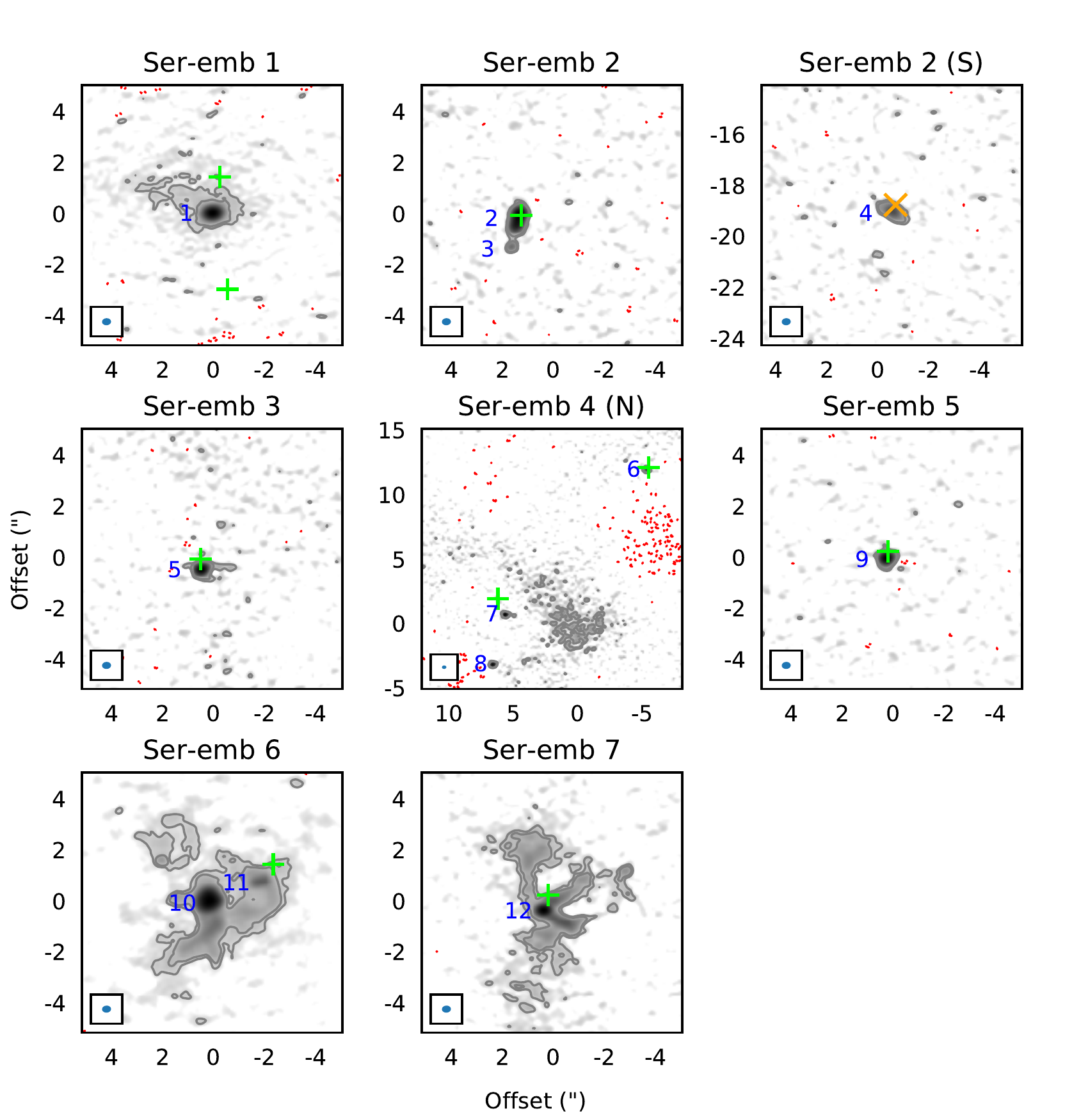}
    \caption{Postage stamps from our ALMA maps of our Serpens Sample of deeply embedded protostars.
        Intensity is shown in negative greyscale with logarithmic scaling to highlight extended
        structure. The x and y axes correspond to the offset from the pointing center (table
        \ref{tab:alma_obs_setup}) of each map. Grey contours are shown at 3 and 5 times the RMS in
        each map, while dashed red contours are shown at -3 times the RMS. Blue numbers indicate the
        ID of the sources in each map which were fit by a Gaussian in table \ref{tab:alma_sources}.
        YSOs previously identified by mid-IR Spitzer \citep{dunham2015} surveys are indicated by
        green pluses (Class 0+I and Flat Spectrum) and orange crosses (Class II and III). Each map
        is shown without primary beam correction for clearer flux scaling. Full maps from each
        pointing are provided in the appendix.}  
    \label{fig:alma_serpens_maps_1}
\end{figure*}

\begin{figure*}[htb] %alma_serpens_maps_2
    \centering
    \includegraphics[scale=1.0]{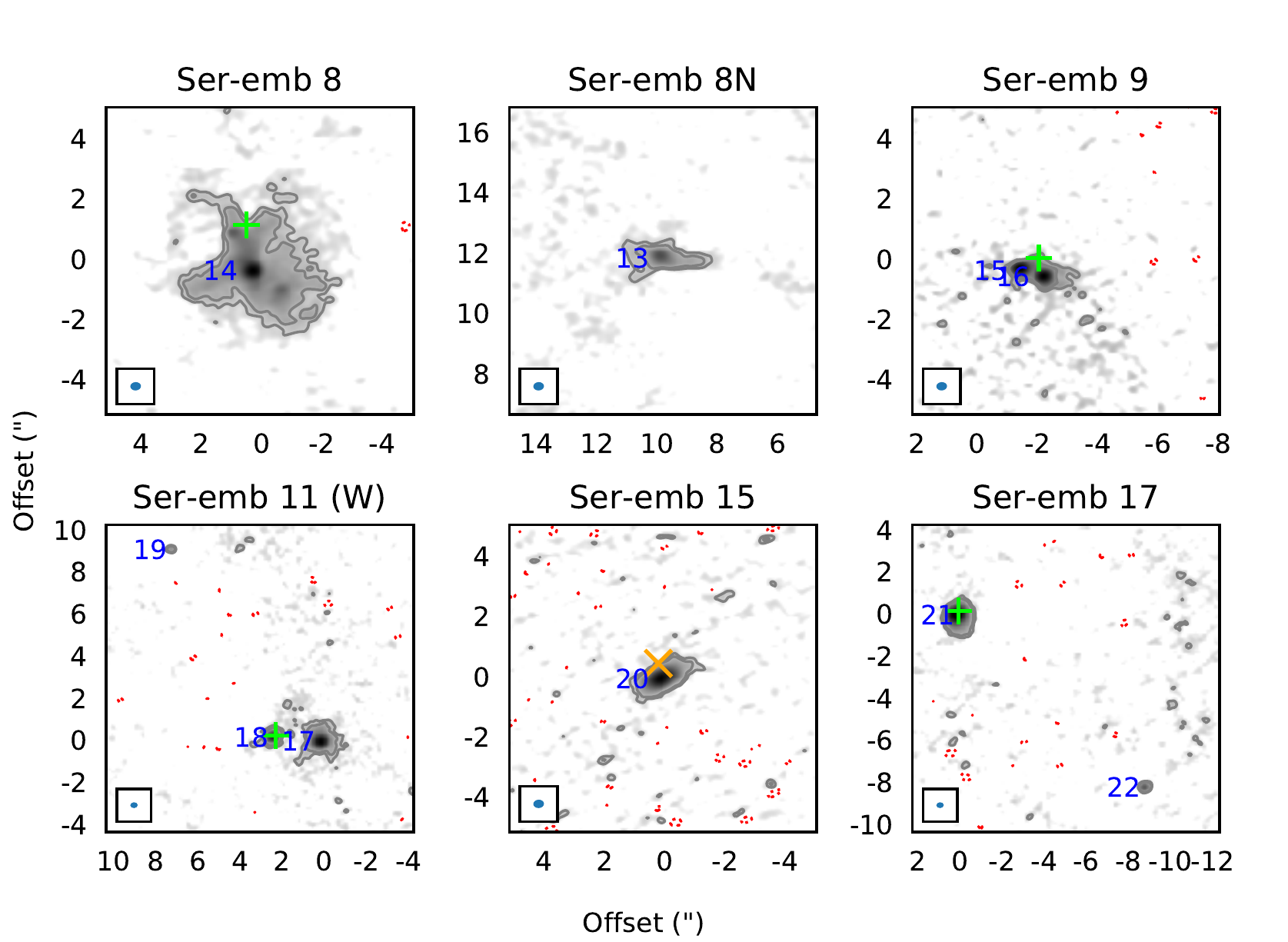}
    \caption{Figure \ref{fig:alma_serpens_maps_1} cont.}
    \label{fig:alma_serpens_maps_2}  
\end{figure*}

Maps of the Serpens protostars targeted by our ALMA observations were produced using the
\texttt{clean} task in CASA 4.7.2. During self-calibration, all channels in each spectral window
were averaged together to increase the SNR. The \texttt{clean} task was then run in multi-frequency
synthesis mode to a threshold of 3$\sigma$ (measured by the RMS in an emission free region of each
image) using Briggs weighting with robust=0.25 and a pixel size of 0.06\arcsec~to produce
60x60\arcsec~maps. All maps were primary beam corrected to a limiting response 
level of 0.1, corresponding to an image diameter of $40\arcsec$. After correction, the flux of 
point-like sources was measured by Gaussian fitting. Postage stamps from the resulting maps are 
shown in figures \ref{fig:alma_serpens_maps_1} and \ref{fig:alma_serpens_maps_2}, while the 
corresponding Gaussian
fits are provided in table \ref{tab:alma_sources}. In each figure, YSOs previously identified by
mid-IR Spitzer surveys \citep{dunham2015} are indicated by green pluses (Class 0, I, and
Flat-Spectrum) and orange crosses (Class II and III).

While \cite{enoch2011} only detected sources towards nine of the twelve Serpens fields surveyed, our
observations find sources in every field owing to ALMA's higher sensitivity (0.1 mJy vs the $> 0.9$ 
mJy in the CARMA maps). Most sources are resolved by the 0.3\arcsec beam, and many are surrounded by
extended structure which may in some cases be evidence of cavity walls sculpted by outflows. As the
ALMA configuration used was selected to filter out spatial scales larger than $3.0$\arcsec, there is
additional extended structure missing from our images. Ser-emb 4 (N) shows a clear example of
this, as it is faint and marginally resolved out by ALMA, but is strongly detected (SNR $> 20$) in
CARMA maps made only with visibilities for scales $> 4.1\arcsec$ \citep{enoch2011}.

In comparisons to the locations of our sources with Spitzer YSOs, there is generally good 
correspondence, however, there are several sources detected in our 1.3 mm maps with no associated 
Spitzer source. Discussion of these sources and further descriptions of each ALMA map are given 
in the appendix.

\section{Detecting Variability in Ideal Comparisons of Interferometer Observations}
%\section{Expectations for ALMA Detections of Variability}
\label{sec:alma_expectations}

The smallest variation in flux of a source that can be robustly detected by comparing two
interferometric observations depends upon the continuum RMS noise, the method used for calibrating
and  measuring flux, and how differences in spatial and spectral configurations are accounted for.
The most ideal situation would be the comparison of observations made with the same telescope and
identical setups, and thus here we compare our 2016 observations against themselves to find
approximate lower limits on detectable flux variations, while leaving discussion of comparing
different interferometer setups to section \ref{sec:alma_to_carma}.

\subsection{Continuum RMS Noise of Observations}

\begin{figure}[htb] %brt_pk_contam
    \centering
    \includegraphics{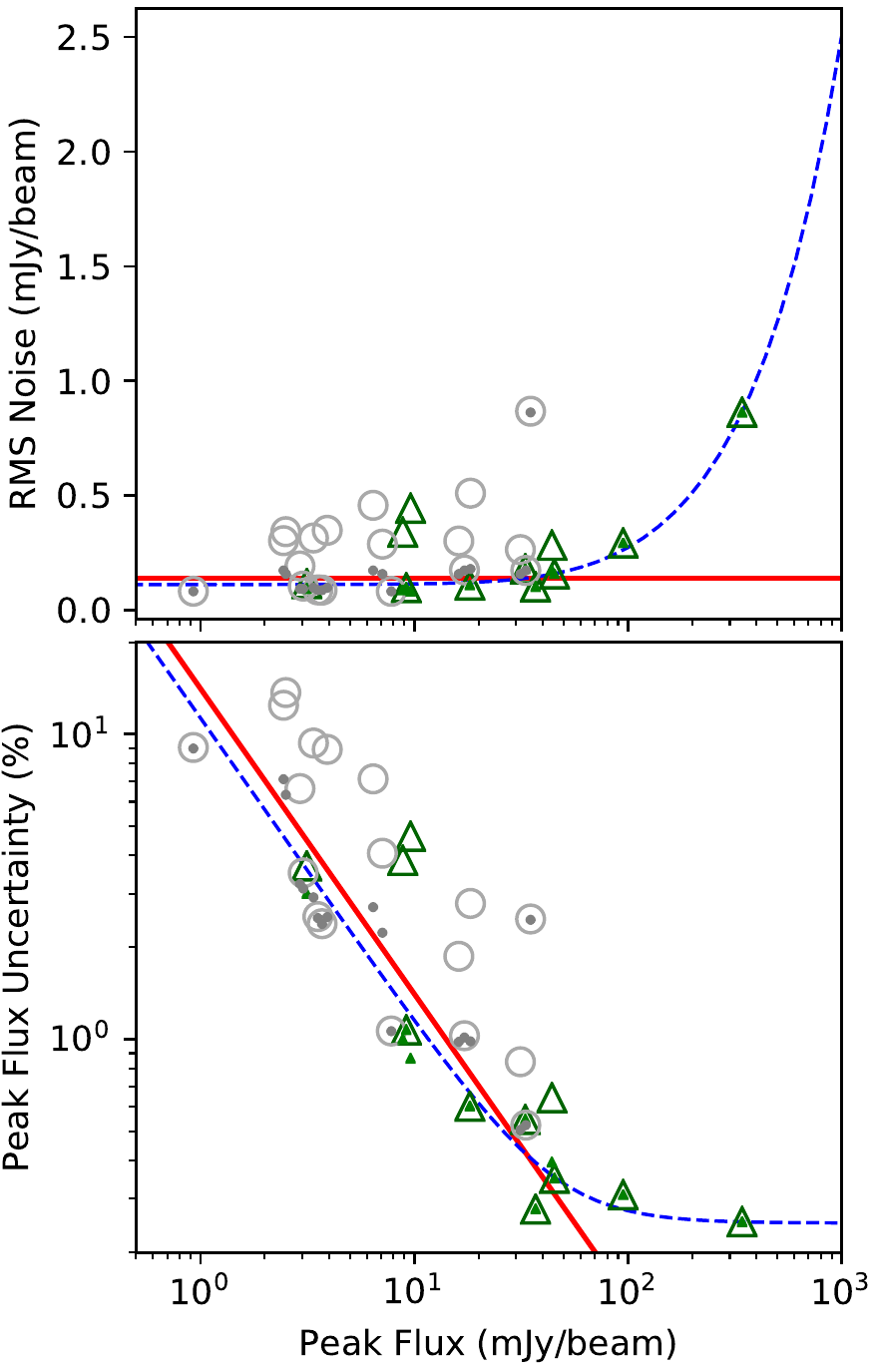}
    \caption{Upper panel: Achieved RMS noise vs the peak flux for each 
        source in the first scan of our ALMA observations. Lower panel: Same as upper panel, but 
        with the \% uncertainty in Peak Flux on the y axis, i.e., the RMS noise divided by the Peak 
        Flux. Large open symbols represent measurements
        accounting for the increased noise due to the reduced sensitivity of the ALMA primary beam
        near the field edge, while small filled symbols use the RMS noise at the field center. Green
        triangles are the brightest sources in a given field, grey circles are from fields with a
        brighter source, which may cause ALMA to reach its dynamic range limit. The red line is
        $\sqrt{2}$ times the requested RMS noise of 0.1 mJy. The blue curve in the upper panel is 
        an empirical fit to the brightest peaks with the RMS noise of the field center of the form 
        $A=\sqrt{R^2+P^2/D^2}$, where $A$ and $R$  are the achieved and requested RMS, $P$ is the 
        peak flux, and $D$ is the fitted dynamic range limit. The blue curve in the lower panel 
        plots this function divided by Peak Flux, $A/P$.}
    \label{fig:brt_pk_contam}
\end{figure}

Our 2016 ALMA observations were requested to reach a continuum RMS noise level of 0.1 mJy, defined
as the RMS in an emission free region of a deconvolved (i.e. cleaned) continuum image. This RMS
noise limit was intended to allow reaching a SNR $>50$ for our targets, where the SNR is defined as
the ratio of peak flux to RMS noise.
\begin{comment}
\footnote{Peak Flux is somewhat of a misnomer, as its units are actually those
    of intensity (e.g. Jy/beam). However, for a point source the flux in a beam is exactly the same 
    as the total flux, and the term ``Peak Flux'' is reasonable.}
\end{comment}
\hspace{-0.5pc}For our ALMA maps produced from the first scan, the achieved RMS noise (and the RMS
as a percentage of the peak flux) is plotted against the peak flux for each source in figure
\ref{fig:brt_pk_contam}. Many of our detected sources have an RMS noise greater than that expected
(i.e., above $\sqrt{2}$ times the red curve), however, this can be readily explained by the reduced
sensitivity of the ALMA dishes to sources near the edge of the field of view and the dynamic range
limit of ALMA. Large open symbols in figure \ref{fig:brt_pk_contam} account for the increased noise
for sources nearer the edge of the field of view, while smaller filled symbols indicate the RMS
noise level had every source been at the field center. Some sources still would lie significantly
above the expected noise level even if they had been observed at the field center (small grey
symbols). Here the noise is dominated by the dynamic range limit of ALMA due to a brighter source in
the same field (green symbols). ALMA's dynamic range limit describes the expected SNR for the
brightest source in the field without self-calibration, and is nominally ~100 for Band 6
observations (ALMA Cycle 3 Proposer's Guide). We find through a fit to the expected noise behaviour
for our observations after self-calibration (blue curve in figure \ref{fig:brt_pk_contam}.) that the
dynamic range limit is $\sim400$. %more exactly, 400 +/-
%1.

\subsection{Comparison of First and Second ALMA Scans} 

To estimate lower limits on detectable flux variations using only ALMA, we divided our visibility
data for each field into its individual scans, then independently self-calibrated and
imaged each scan using the same \texttt{CASA clean} parameters as those for the full data set in
section \ref{sec:alma_maps}. Integrated and peak fluxes were measured for each source by an
elliptical Gaussian fit using \texttt{CASA imfit}. We also measured the integrated and peak flux in
fixed regions of the sky enclosing each source (``Box Method''), typically a square 1-1.5\arcsec~ 
in size.
For this method, the uncertainty in the peak flux is the RMS noise, while that for the integrated
flux is $\sqrt{N}$ times the RMS noise, where $N$ is the number of pixels in the region. The  use of
the Box Method for measuring flux is motivated by the large number of sources which are resolved
and/or embedded in extended structure, and therefore not well described by a Gaussian model.

Regardless of how flux measurements are made on the images, direct comparisons between two ALMA
observations will be limited by the nominal Band 6 flux calibration accuracy of $\sim$10\% 
\citep{almatechcycle3}. Poor flux calibration accuracy is a general issue with mm/sub-mm
observing caused by the paucity of bright, stable point sources
\footnote{mm/sub-mm observations are most often calibrated using 
    bright quasars, or if available, solar system planets. Unfortunately, Quasars are highly 
    variable at these wavelengths, while planets are typically resolved and require very 
    accurate flux models.}.  
        
\begin{figure*}[htb]
    %"Agreement" is the difference between values of each measure divided by the error in the 
    %difference, which is the error of each measurement added in quadrature.
    \includegraphics{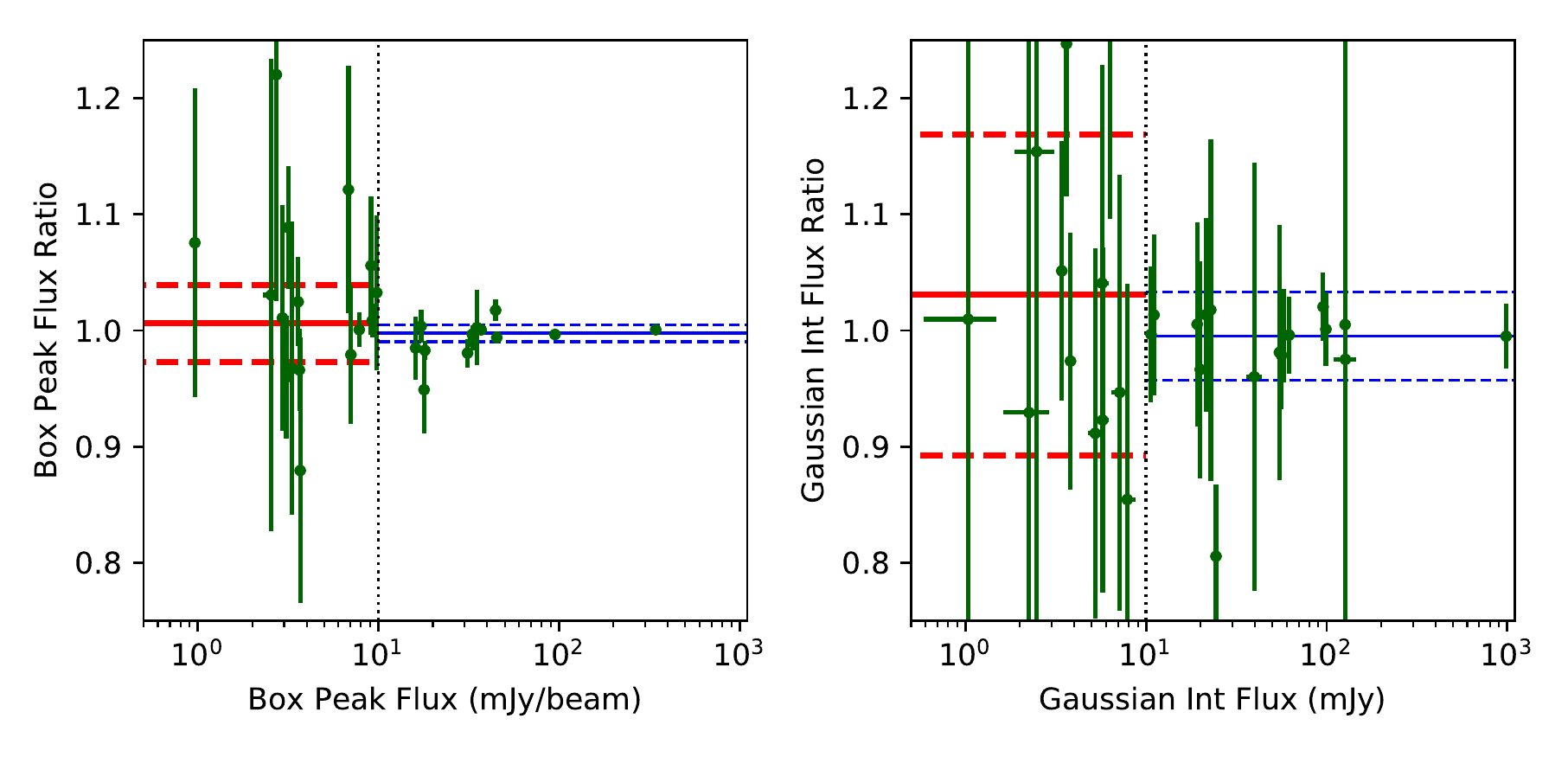}
    \caption{Ratio between the second and first scans in peak flux 
        estimated using a fixed region (box) on the sky and integrated flux estimated by a Gaussian 
        fit. The average weighted 
        by $\sigma^{-2}$ and the associated standard deviation are shown by the solid and dashed 
        lines for dim sources ($<10$ mJy or mJy/beam ), and bright sources ($>10$ mJy or mJy/beam). 
        The values of these averages are summarized in table \ref{tab:flux_comp_fcfs}}.
    % Note that Ser-emb 5 was observed for three scans, and has therefore been omitted from 
    %this comparison.}
    \label{fig:flux_comp_rat}
\end{figure*}

To sidestep this problem, we turn to relative flux calibration methods similar to those used in the
JCMT Transient Survey \citep{mairs2017a}, and apply them to both our predictions here and our
comparison of ALMA and CARMA observations in section \ref{sec:alma_to_carma}. We determine Relative
Flux Calibration Factors (rFCFs) to bring the flux scale of the first scan into agreement with the
second by fitting an average to the flux ratios between scans for bright sources 
($>10$ mJy)\footnote{At the requested $100$~$\mu$Jy RMS noise of our 
observations, the bright sources
    are those for which we can achieve the desired SNR $>$ 100.}. We also separately find rFCFs for 
    only
the dim sources ($<10$ mJy or mJy/beam ) as a sanity check and to see what level of variability 
could be detected without bright sources. When fitting the average to determine the rFCF, each 
point is
weighted by $\sigma^{-2}$, where $\sigma$ is the uncertainty in the ratio given by the errors added
in quadrature of the flux measurements in the first and second scan. The overall uncertainty in the
rFCF is given by the standard deviation of the ratios, again weighted by $\sigma^{-2}$. The ratios
and rFCF fits for just the Box peak flux and Gaussian integrated flux are shown in figure
\ref{fig:flux_comp_rat}, while table \ref{tab:flux_comp_fcfs} summarizes all rFCF values. All of the
rFCFs are consistent with 1, and most deviate by $\lesssim 0.01$, demonstrating that the ALMA
calibration is extremely stable between scans on the $\sim40$ minute timescale of our observations.
The precision of rFCFs derived from integrated fluxes are typically lower than those derived from
peak fluxes by a factor of 2-5, due to the larger relative uncertainties in integrated flux. The
precision for rFCFs determined using Gaussian fits are lower than those using the peak/integrated
Box flux because we assume flux measurements are independent between scans, yet many of our sources
are poorly described by a Gaussians and thus have flux uncertainties dominated by the quality of
fit. This causes the flux uncertainties to be correlated scan-to-scan, resulting in a larger rFCF
uncertainty. The best rFCF precision is thus achieved using the Box peak flux (essentially the SNR),
with a precision of 0.7\% and 3.3\% for bright and dim sources respectively. This is consistent with
what would be expected from the inverse of the SNR for representative dim ($\sim30$) and bright
($\gtrsim100$) sources.

\begin{deluxetable}{lcc} %flux_comp_fcfs
    \tablewidth{0pt}
    \tablecolumns{3}
    \tablecaption{Relative Flux Calibration Factors (rFCFs) for ALMA data}
    \tablehead{rFCF Measurement Method & Dim & Bright }
    \startdata
    Box Peak Flux & 1.006 (0.033) & 0.998 (0.007)  \\
    Box Int Flux & 0.990 (0.152) & 0.997 (0.012) \\
    Gaussian Peak Flux & 1.012 (0.074) & 1.002 (0.012) \\
    Gaussian Int Flux & 1.031 (0.138) & 0.995 (0.038) 
    %\tablecomments{}
    %\tablenotetext{a}}
    \enddata
    \label{tab:flux_comp_fcfs}
\end{deluxetable}

It should be emphasized that reaching sensitivity to low levels of variability requires both a 
precisely determined rFCF \emph{and} high SNR flux measurement for an individual source. Table 
\ref{tab:var_thresholds} summarizes the percentage change in flux we would be sensitive to at a 
$3\times$RMS level for a given rFCF and Flux percentage error (assuming that 
the Flux percentage 
error 
does not change between observations). Using the Box peak flux rFCFs, we are thus sensitive to 
variability at the $\sim16\%$ level for representative dim sources (3 mJy; 3\% flux error) 
and at the $\sim4.8\%$ level for bright sources (10 mJy; 1\% flux error).

Furthermore, it should be noted that since our observations were taken in a 
snapshot mode, the uv-coverage of each scan used in these comparisons is nearly identical. To 
estimate how small differences in uv-plane sampling might affect our sensitivity to variability, we 
conducted additional tests where a distinct subset of antennas was dropped at random from each 
scan, and rFCFs were calculated using the Box peak flux as described above. These tests were done 
removing 2, 4, or 8 of the 39 antennas from each scan, which is equivalent to having 89\%, 78\%, or 
55\% of visibilities from each scan at identical positions in the uv-plane. The resulting rFCF 
values remain consistent with 1.0, while the sensitivity to variability for a 10 mJy source with 2, 
4, and 8 antennas randomly dropped respectively becomes 4.8\% (unchanged), 5.6\%, and 7.8\%.  This 
suggests that small differences in uv-plane sampling will not significantly affect our ability to 
detect variability. The general impact of larger differences in uv-plane sampling is discussed in 
depth in section 5.
   
\begin{deluxetable}{c|lllllll} %var_thresholds
    \centering
    \tablecaption{3$\times$RMS Percentage Variability detection Thresholds}
    \tablehead{\multirow{2}{*}{Flux Err. (\%)} & \multicolumn{7}{c}{rFCF Err. (\%)} \\
        & \colhead{0.3} & \colhead{0.5} & \colhead{1} & 
        \colhead{3} & \colhead{5} & \colhead{10} & \colhead{20}}
    \startdata
    0.3 & 1.6 & 2.0 & 3.3 & 9.1 & 15.1 & 30.0 & 60.0 \\
    0.5 & 2.3 & 2.6 & 3.7 & 9.2 & 15.1 & 30.1 & 60.0 \\
    1 & 4.3 & 4.5 & 5.2 & 9.9 & 15.6 & 30.3 & 60.1 \\
    3 & 12.8 & 12.8 & 13.1 & 15.6 & 19.7 & 32.6 & 61.3 \\
    5 & 21.2 & 21.3 & 21.4 & 23.0 & 26.0 & 36.7 & 63.6 \\
    10 & 42.4 & 42.5 & 42.5 & 43.4 & 45.0 & 52.0 & 73.5 \\
    20 & 84.9 & 84.9 & 84.9 & 85.3 & 86.2 & 90.0 & 103.9
    \enddata
    \label{tab:var_thresholds}
\end{deluxetable}
\hspace{1pc}

\section{Detecting Variability between Distinct Interferometric Observations with ALMA and CARMA}
\label{sec:alma_to_carma}

While the results of section \ref{sec:alma_expectations} provide a lower limit for detections of
variability in ideal conditions, they do not take into account many of the complications involved
in comparing typical interferometric observations. These issues are caused by the inherent
flexibility of interferometers, which typically have multiple array configurations for recovering
structure over a range of spatial scales, and a variety of frequency bands and correlator modes for
sampling different parts of the spectrum.

 For comparing our ALMA and CARMA observations, we first attempt to address the
problems caused by differences in array configuration in sections
\ref{ssec:spatial_and_spectral_diff}-\ref{ssec:sim_obs}, and then turn to relative flux calibration
methods for searching for variability in \ref{ssec:relative_fcfs}.

\subsection{Impact of Differences in Spatial Configurations}
\label{ssec:spatial_and_spectral_diff}

Interferometers sample the complex visibility $V(u,v)$ of a source, a Fourier transform of it's
intensity distribution on the sky (ignoring effects of the primary beam) and a function of the
spatial frequencies $u$ and $v$. The spatial frequencies sampled are determined by the projected
lengths of the array's baselines (in units of the observing $\lambda$) on the sky in the East-West
($u$) and North-South ($v$) directions. Since the projected baseline lengths and orientations change
as the earth rotates, the $uv$-plane becomes better sampled over the course of observation, however,
this implies that no two interferometric observations will recover exactly the same visibilities
unless they observe with identical array configurations and observing schedules, from the same
latitude, and at the same wavelength. Since the synthesized beam is simply the Fourier transform of
the (weighted) visibility sampling function, this is equivalent to stating that two observations
will not have the same beam unless subject to the above conditions.

While differences in $uv$-plane sampling are not a problem for imaging point sources, (which have a
constant visibility amplitude over the $uv$-plane) images of resolved sources constructed by 
inverting $V(u,v)$ will
contain varying amounts of flux depending on the $uv$-plane sampling. This issue is partly mitigated
by algorithms used in imaging such as CLEAN, which effectively estimate $V(u,v)$ in un-sampled
regions of the uv-plane by interpolating between samples, however, extrapolating to regions with no
samples at all is extremely difficult. In particular, CLEAN has trouble in accurately extrapolating
to the center of the $uv$-plane (where large spatial scales are measured), which is typically poorly
sampled by interferometers because of a lack of short baselines.

An example is helpful in further illustrating how differences in $uv$-plane sampling can affect the
recovered flux. Consider a point source embedded in extended structure observed by two different
configurations of the same interferometer - the first with the antennas  in a group compact enough
to recover the largest spatial scales of the extended structure, and the second with the antennas
spaced further apart, providing higher resolution but missing some of the extended structure. If
images are produced for both observations and the flux of the point source is to be measured, the
compact configuration image can be used to make a more accurate estimate of the point source flux by
fitting for both the point source and the underlying larger scale structure.

To account for this bias, changes in the flux of this point source between the two observations
should be measured using images constructed only using visibilities which measure similar spatial
scales. There are several ways to accomplish this which we discuss in the following sections,
including $uv$-plane matching of synthesized beams and simulated observations.

\subsection{$uv$-plane Matching of ALMA and CARMA Synthesized Beams}
\label{ssec:beam_matching} 

\begin{figure}[htb] %beam_matching_demo
    %\centering
    \hspace{1pc}
    \includegraphics[scale=1.0]{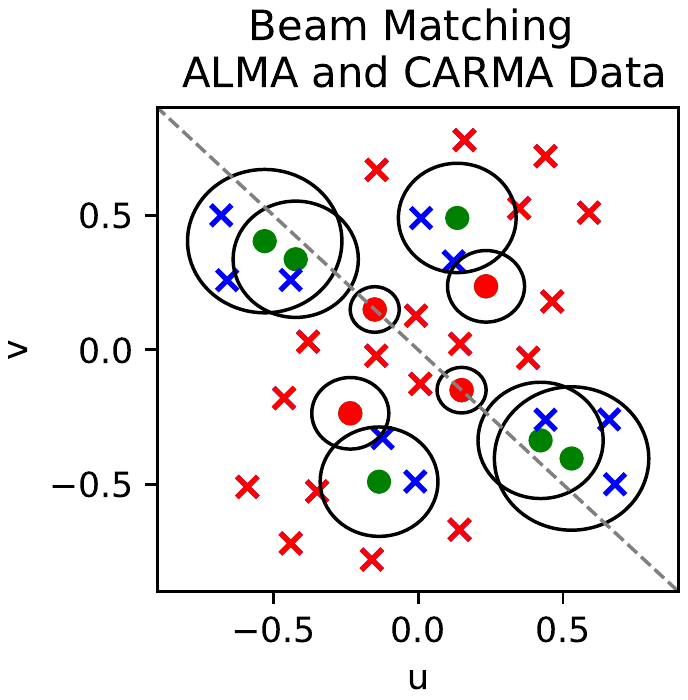}
    \caption{Demonstration of how beams are matched for our observations using two
        fictitious data sets. ALMA and CARMA visibilities in the $uv$-plane are indicated by 
        blue/red crosses and green/red disks respectively. Red disks and crosses are visibilities 
        which will be removed from each data set by beam matching. The dashed grey line shows the 
        inherent symmetry axis of the $uv$-plane. The black circles around each CARMA visibility 
        indicate the cut-off distance for beam matching; if there are no samples from the ALMA 
        data set within the cut-off distance (here, $f_\text{cut}=0.4$), the CARMA visibility in 
        the original data set is removed. Any ALMA visibilities which do not fall within the 
        cut-off distance to a CARMA visibility are also removed. }
    \label{fig:beam_matching_demo}
\end{figure}

To directly compare our ALMA and CARMA data for a given source, we first include only visibilities
from each observation which sample a similar region of the $uv$-plane, in order to match the
synthesized  beam shapes as closely as possible. This requires some care, as a simple euclidean
distance is not appropriate for comparing $uv$-plane separations between visibilities. The angular
scale a visibility measures is the inverse of its euclidean distance from the $uv$-plane origin, and
thus a given euclidean distance between two points near the origin is equivalent to a much larger
change in angular scale than for the same distance between two points far from the origin. To avoid
this problem, we instead use the euclidean distance between ALMA and CARMA visibilities as a
\emph{fraction} of the $uv$-distance to the CARMA visibility from the origin. Figure
\ref{fig:beam_matching_demo} shows in detail  how this fractional distance is used for matching two
fictional ALMA and CARMA data sets with an unrealistically large cutoff distance of
$f_\text{cut}=0.4$.

\begin{figure*}[htb] %beam_matching_real
    \centering
    \includegraphics[scale=1.0]{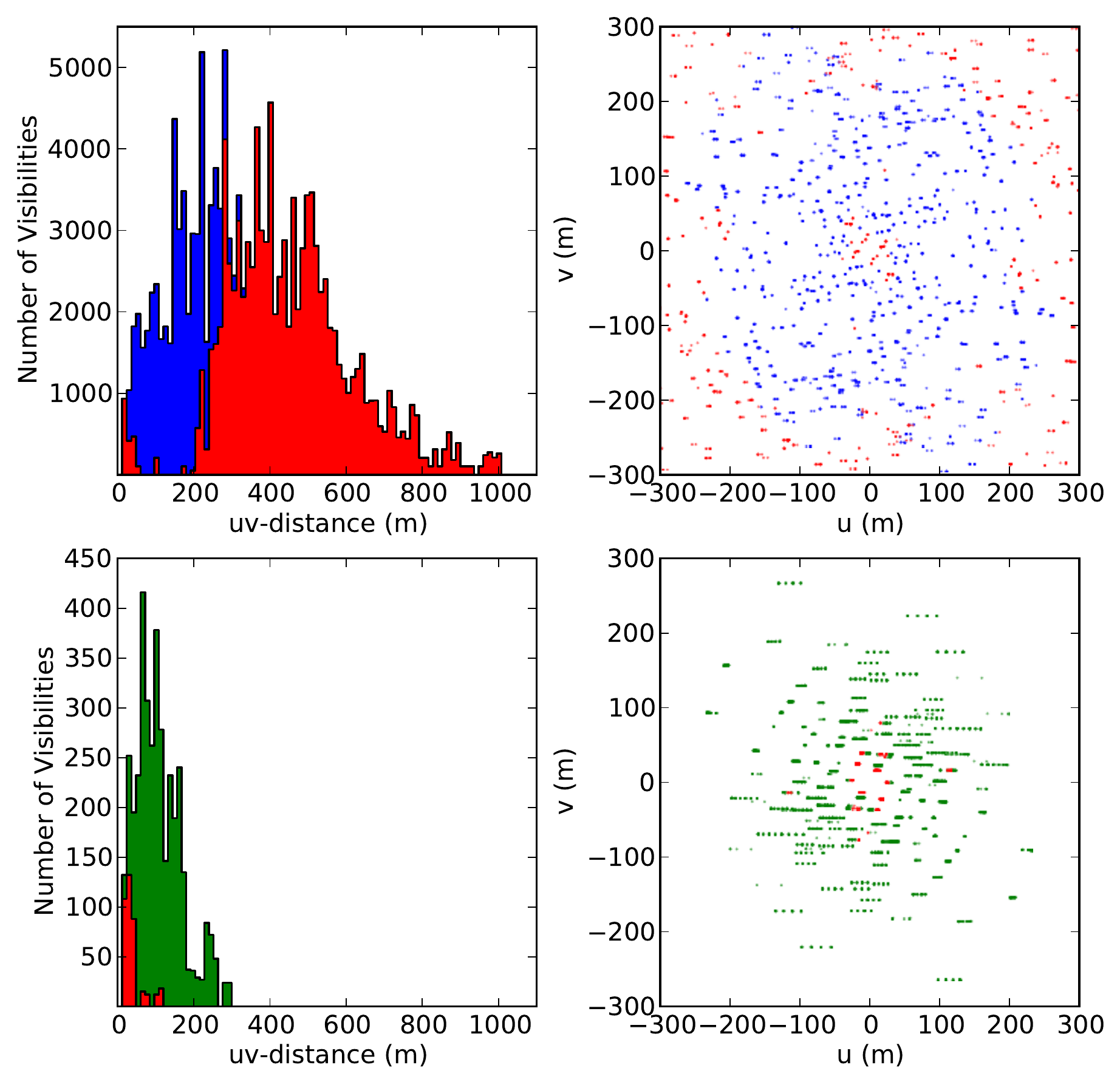}
    \caption{Effect of applying beam matching with $f_\text{cut}=0.25$ on the $uv$-plane 
    distributions of visibilities for CARMA Track C1.2 and our ALMA observations of Ser-emb 1. The 
    top and bottom rows of panels shows the ALMA and CARMA distributions respectively, where red 
    indicates visibilities removed by beam matching.}
    \label{fig:beam_matching_real}
\end{figure*}

\begin{figure}[htb] %alma_carma_nn_kde
    \centering
    \includegraphics[scale=1.0]{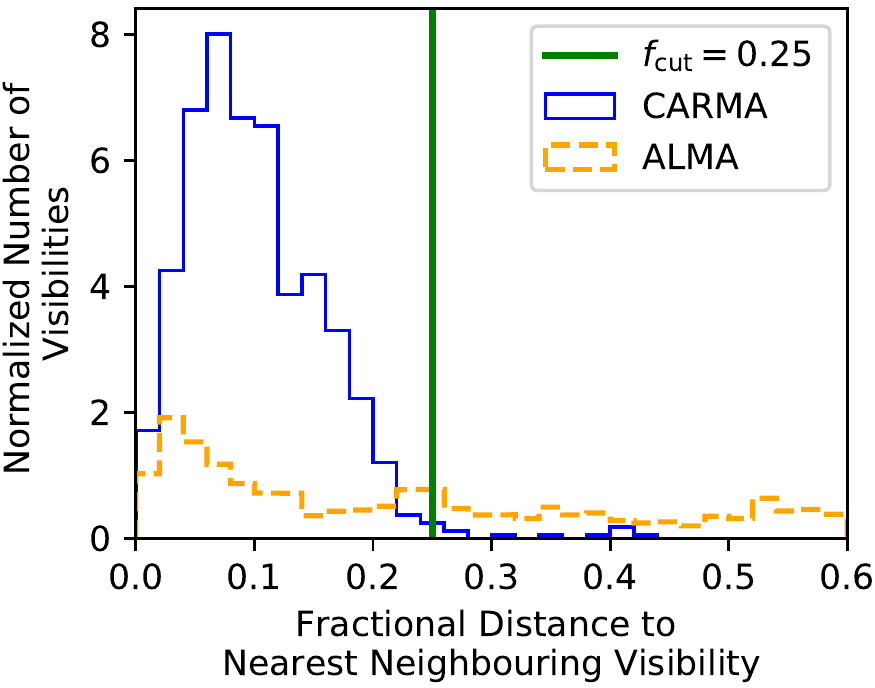}
    \caption{Normalized histograms of nearest neighbouring visibilities (in the 
    other data set) for our ALMA and the CARMA track C1.5 observations of Ser-emb 6. The bin size 
    is 0.02. The fractional distance cut-off $f_\text{cut}$ used for beam matching of all ALMA and 
    CARMA observations is shown by the vertical line.}
    \label{fig:alma_carma_nn_kde}
\end{figure}

For comparisons of our observations, we wish to optimize the value of $f_\text{cut}$ by choosing it
to match the beam dimensions as closely as possible (smaller $f_\text{cut}$) without removing so
much data that the SNR drops severely (larger $f_\text{cut}$). Since the distribution of baselines
for our ALMA configuration is essentially a superset of those for CARMA C (see figure
\ref{fig:bl_comp_kde}), a useful value of $f_\text{cut}$ should mostly remove visibilities from the
ALMA data while retaining as many from the (much lower SNR) CARMA data as possible. Figure
\ref{fig:alma_carma_nn_kde} shows the distributions of nearest neighbours between a pair of ALMA and
CARMA observations of the same Serpens source in units of fractional distance, from which a value of
$f_\text{cut}$ of $\sim0.25$ appears optimal. Since the $uv$-coverage is similar for all other
pairs of observations, we use $f_\text{cut}=0.25$ in every comparison. Figure 
\ref{fig:beam_matching_real} shows an example of the $uv$-plane distributions before and after 
applying beam matching with $f_\text{cut}=0.25$ to a pair of ALMA and CARMA observations of the 
same source.

Once the visibility data from each observation to be compared are selected, preliminary maps are
produced with the same parameters as we did for the full ALMA maps in section \ref{sec:alma_maps}.
In table \ref{tab:beam_matching_shapes}, we compare the beam shapes before and after matching. While
correspondence in the ALMA and CARMA synthesized beams has improved with their respective increase
and decrease in size, CARMA's beam is still typically $\sim0.3\arcsec$ larger in either axis than
ALMA's. This can be attributed to differences in the $uv$-plane sampling density and weighting of
the visibilities which $uv$-plane matching does not take into account. The sampling density of ALMA
is significantly better than CARMA at all $uv$-distances, but is relatively skewed towards large
$uv$-distance owing to the longer baselines. Furthermore, longer baseline tend to also have larger 
phase scatter from atmospheric fluctuations, resulting in a drop in visibility amplitude 
\citep{zauderer2016}. These noisier visibilities are flagged or given a lower weight during 
calibration, resulting in less sensitivity on longer baselines. This affects the data from CARMA 
more than that from ALMA because of the poorer observing conditions at the CARMA site. 
To correct for these effects, we also apply a $uv$-taper to each ALMA observation when
imaging the data (last column of table \ref{tab:beam_matching_shapes}). The taper used is equivalent
to smoothing by a circular Gaussian kernel with FWHM equal to the major axis of the corresponding
matched CARMA beam, and improves the agreement in beam shapes to within $\sim0.1\arcsec$ or better
for both axes.

\begin{deluxetable*}{lcccccc} %beam_matching_shapes
    \tablewidth{0pt}
    \tablecolumns{7}
    \tablecaption{Beam Shapes Before and After $uv$-plane Matching}
    \tablehead{\multirow{2}{*}{Matching Track\tablenotemark{a}} & \multicolumn{2}{c}{Unmatched 
    (\arcsec)} & 
    &  
    \multicolumn{3}{c}{Matched (\arcsec)}\\
        \colhead{} & \colhead{CARMA} & \colhead{ALMA} & & \colhead{CARMA} & 
        \colhead{ALMA} & \colhead{ALMA + $uv$ taper}} 
    \startdata
    C1.2 & 1.73 x 1.40 &\multirow{4}{*}{0.35 x 0.27} &\hspace{2pc} & 1.16 x 1.00 & 0.82 x 0.65 & 
    1.19 x 1.02 
    \\    
    C1.5 & 1.72 x 1.62 &                             &\hspace{2pc} & 1.46 x 1.21 & 1.12 x 0.95 & 
    1.53 x 1.38 
    \\     
    C1.8 & 1.67 x 1.43 &                             &\hspace{2pc} & 1.30 x 0.85 & 0.92 x 0.54 & 
    1.32 x 1.10 \\ 
    C2.3 & 1.54 x 1.39 &                             &\hspace{2pc} & 1.02 x 0.82 & 0.69 x 0.49 & 
    1.04 x 0.87 
    \enddata
    \tablecomments{Beam sizes given are the FWHM of the major and minor axes.}   
    \tablenotetext{a}{ Since the $uv$-coverage for each source within an observation is nearly 
    identical, we only compare the beams for one source observed by both.}
    \label{tab:beam_matching_shapes}
\end{deluxetable*}

\subsection{Simulated Re-observations of ALMA sources with CARMA}
\label{ssec:sim_obs}

Simulated observations provide another means of comparing our ALMA and CARMA data while taking into
account differences in $uv$-plane sampling. Typically, simulated observations are used to predict
what an interferometer will see with a  particular array configuration and observing setup given a
model of the true sky brightness. The model is often the result of a numerical simulation or an
image from another telescope. In our case, we use the maps derived from our ALMA observations in
section \ref{sec:alma_maps} with the residuals subtracted as models, and then use CASA's
\texttt{simobserve} task to simulate what CARMA would see had it re-observed the same sources in a
identical manner as it did in 2007, and compare against the beam matched CARMA maps from section
\ref{sec:alma_to_carma}.2. A similar approach was successfully used by \cite{hunter2017} for
detecting variability between SMA and ALMA observations of the massive protocluster NGC 6334-I, 
albeit with a readily detectable factor of $\sim4$ change in flux.

In order to realistically simulate the conditions under which the original CARMA data were taken,
the same array configuration, correlator setup, observing schedule, and data flagging must be used.
For our simulations, we ensure the $uv$-coverage is the same by extracting the exact positions of
the CARMA antennas from the data and applying the same observing schedule, integration time, and
flagging. For the spectral setup,  we simply use a single 2.8 GHz wide spectral window centred on
the ALMA observing frequency (227 GHz), rather than the exact parameters in table
\ref{tab:carma_c_obs_setup} due to limitations of the CASA \texttt{simobserve} task. While the
observing frequencies of ALMA and CARMA are slightly different, changes in the flux should be small 
and mitigated by relative flux calibration (see discussion in section \ref{sec:dis_and_conc}). 
Finally, we do not add any noise to the simulated visibilities, as we are only modelling 
differences in $uv$-coverage that would affect measurement of variability, and not attempting to 
determine sensitivities for a CARMA to CARMA comparison.

From the simulated visibilities, maps are produced using the same parameters as those in section
\ref{sec:alma_maps}, but only including visibilities more than 30m from the $uv$-plane origin. We 
apply this $uv$-plane cut as a good approximation to the effects of beam matching on the CARMA 
data\footnote{This also ensures that the simulated CARMA maps do not contain larger spatial 
scales than the ALMA observations used as models (with relatively fewer short baselines than the 
CARMA C configuration) were sensitive to.}. 
The simulated and $uv$-plane matched beam shapes are compared against each
other in table \ref{tab:sim_obs_beam_shapes}. As was similarly the case for beam matching,
differences in beam shape of $\sim 0.2\arcsec$ remain, however, this likely arises from variations
in the visibility weighting not modelled in the simulations, as the $uv$-coverages of the beam 
matched and simulated observations are nearly identical. To correct for this, a $uv$-taper 
corresponding to the CARMA matched beam is added to each simulated image when cleaning as was done 
for beam matching.
    
\begin{deluxetable}{lccc} %sim_obs_beam_shapes
    \tablewidth{0pt}
    \tablecolumns{4}
    \tablecaption{CARMA Beam Shapes In Real and Simulated Observations}
    \tablehead{\colhead{Track} & \colhead{Matched (\arcsec)} & \colhead{Simulated (\arcsec)} & 
    \colhead{Sim. + Taper (\arcsec)}}
    \startdata
    C1.2 & 1.16 x 1.00 & 1.08 x 0.71 & 1.18 x 0.95 \\
    C1.5 & 1.46 x 1.21 & 1.23 x 1.06 & 1.44 x 1.25 \\ 
    C1.8 & 1.30 x 0.85 & 1.09 x 0.75 & 1.24 x 1.05\\
    C2.3 & 1.02 x 0.82 & 0.89 x 0.77 & 1.02 x 0.92
    \enddata
    \tablecomments{Beam sizes given are the FWHM of the major and minor axes.}  
    %\tablecomments{a}
    %\tablenotetext{a}}
    \enddata
    \label{tab:sim_obs_beam_shapes}
\end{deluxetable}

\subsection{Relative Flux Calibration Factors and Variability of Sources}
\label{ssec:relative_fcfs}

With differences in beam shapes minimized, we can reliably measure fluxes of sources common to both
observations and attempt searches for variability. As discussed in section
\ref{sec:alma_expectations}, direct comparisons are limited by the accuracy of the absolute
flux calibration for each telescope, $\sim$10\% for ALMA in Band 6 (ALMA Cycle 3 Technical Handbook)
and $\sim$20\% for the CARMA observations \citep{enoch2011}. We therefore 
calculate rFCFs by fitting an average
to the ratio of ALMA to CARMA Box peak fluxes, and use them to convert the CARMA maps to the ALMA
flux scale for both methods of beam comparison. Specifically, rFCFs for $uv$-plane beam matching
(section \ref{sec:alma_to_carma}.2) are fit to the ratio of fluxes in pairs of beam matched CARMA 
and ALMA maps (specific to each CARMA track), while rFCFs for simulated re-observation (section
\ref{sec:alma_to_carma}.3) are fit to the ratio of fluxes between the ALMA maps with simulated
re-observation by CARMA and the corresponding beam matched CARMA track. The resulting rFCFs for
CARMA tracks C1.2, C1.5, and C2.3 and the standard deviation between them, $\sigma_\text{rFCF}$, are
shown in table \ref{tab:fcf_comp}. Within the uncertainties given, differences in the flux 
calibration factors across beam comparison methods are consistent with each other and reach a 
similar level of precision. Further comparison of the two methods and discussion of which may be 
best suited to future variability studies is given in section \ref{sec:dis_and_conc}. 

In table \ref{tab:fcf_comp}, a rFCF can not be properly calculated for CARMA track C1.8, as there is
only a single observable source in common with the ALMA observations. Relative variability can thus
not be detected, and the rFCF given is just the ratio of the brightness of this source (Ser-emb 1,
ID 1). This ratio for C1.8 is consistent with the $\sigma_\text{rFCF}$ found for the other three
tracks. 

Given that there is only one ALMA flux calibration for every image, $\sigma_\text{rFCF}$ is
effectively an estimate of the CARMA absolute flux calibration accuracy. We find
$\sigma_\text{rFCF}$ to be $\sim40\%$ for either beam comparison method compared to the $\sim15\%$
expected from the nominal CARMA absolute flux accuracy. Some of this loss of accuracy is probably
caused by the low SNR of the sources used for relative calibration, and some might be attributed to
poorer-than-average weather during several of the CARMA tracks. Since $\sigma_\text{rFCF}$ was only
determined from 3 tracks however, it is not particularly robust.

For tracks C1.2, C1.5, and C2.3, the rFCFs may be useful for measuring variability, but are still
not particularly well constrained due to there only being 3-4 sources per track in common with our
ALMA observations (see table \ref{tab:carma_tracks_sources}). Determining accurate rFCFs is further
hampered by the low signal to noise detections of many of the sources in the CARMA tracks, with most
detected with a SNR of 5-15 and the brightest at a SNR of 30. For tracks C1.2 and C1.5, the relative
flux calibration is good to the $\sim10\%$ level, while for the fainter sources compared in C2.3,
the relative flux calibration is accurate to $\sim50\%$, no better than the $\sim18\%$ uncertainty
due to the combined absolute flux calibration of ALMA and CARMA. 
    
Tables \ref{tab:var_srcs_1}-\ref{tab:var_srcs_3} compare the fluxes in each epoch for each track
after application of the rFCFs. Here, the detection $\sigma$ is the the percent difference between
the fluxes divided by it's uncertainty. No source is seen to vary above $\gtrsim 1 \sigma$ by either
the beam matching or simulated observation analysis. The uncertainty in the percent difference for 
each source implies our ability to detect variability at a 3$\sigma$ level is $\sim24-60\%$ for 
tracks C1.2 and C1.5 and $\sim75-180\%$ for track C2.3, depending on the brightness of the source 
in question. These detection thresholds are much larger than those expected from comparison of 
multiple epochs of ALMA data in section \ref{sec:alma_expectations}, emphasizing the need for large 
numbers of bright sources to achieve good sensitivity to variability.

\begin{deluxetable*}{llcccc} %fcf_comp, unweighted
    \tablewidth{0pt}
    \tablecolumns{6}
    \tablecaption{Relative Flux Calibration Factors}
    \tablehead{\multirow{2}{*}{Beam Comparison Method}  
        &\multicolumn{4}{c}{CARMA Track Name} & \multirow{2}{*}{$\sigma_\text{rFCF}$}\\  
        & \colhead{C1.2} & \colhead{C1.5} & \colhead{C1.8\tablenotemark{a}} & \colhead{C2.3}}
    \startdata
    Beam Matching            & 1.32 (0.09) & 0.74 (0.11) & 1.02 (0.06) & 1.85 (0.54) & 0.45 \\
    Simulated Re-Observation & 1.15 (0.14) & 0.64 (0.04) & 0.93 (0.05) & 1.56 (0.49) & 0.37 
    \enddata 
    \tablenotetext{a}{C1.8 Only has one source, Ser-emb 1 (ID 1), and the uncertainty of the rFCF 
        has been replaced by the uncertainty in the ratio of the fluxes for this track. It is not 
        included in the estimate for {$\sigma_\text{rFCF}$}}
    \label{tab:fcf_comp}
\end{deluxetable*}

\begin{deluxetable*}{llccccccccc} %var_srcs_1
    \centering
    \tablewidth{0pt}
    \tablecolumns{10}
    \tablecaption{Variability of Sources, CARMA Track C1.2}
    \tablehead{\multirow{2}{*}{Beam Comparison Method}& \multirow{2}{*}{ID} &
        \colhead{Beam Matched \& Scaled} & \multirow{2}{*}{Equivalent ALMA Flux} & 
        \multirow{2}{*}{Percent Difference} & \multirow{2}{*}{Detection $\sigma$} \\
         & &CARMA Flux\tablenotemark{a}& & & \\
         & & (mJy beam$^{-1}$) & (mJy beam$^{-1}$) & & }
    \startdata
    \multirow{3}{*}{$uv$-plane Beam Matching}
        & 1   & 114.07 (16.31)  & 125.33 (0.69)  & 9.88 (15.72)   & 0.63 \\
        & 10  & 876.38 (67.20)  & 847.56 (5.47)  & -3.29 (7.44)   & 0.44 \\
        & 11  & 121.95 (26.22)  & 113.91 (5.48)  & -6.59 (20.58)  & 0.32 \\\hline\\
    \multirow{3}{*}{Simulated Re-Observation\tablenotemark{b}}
        & 1   & 99.44 (17.58)   & 114.10 (0.02)  & 14.75 (20.28)  & 0.73 \\
        & 10  & 763.97 (98.70)  & 773.80 (0.21)  & 1.29 (13.09)   & 0.10 \\
        & 11  & 106.31 (25.39)  & 89.26 (0.21)   & -16.03 (20.05) & 0.80
    \enddata 
    \tablenotetext{a}{The flux for $uv$-plane Beam Matching and Simulated Re-observations differ 
    only in the 
    rFCF from table \ref{tab:fcf_comp}.}
    \tablenotetext{b}{The uncertainty in the Equivalent ALMA flux measurements with simulated 
    re-observations is lower than in beam matching because the simulations do not include the 
    effects of noise; see section \ref{sec:alma_to_carma}.2}. 
    \label{tab:var_srcs_1}
\end{deluxetable*}

\begin{deluxetable*}{llccccccccc} %var_srcs_2
    \centering
    \tablewidth{0pt}
    \tablecolumns{10}
    \tablecaption{Variability of Sources, CARMA Track C1.5}
    \tablehead{\multirow{2}{*}{Beam Comparison Method}& \multirow{2}{*}{ID} &
        \colhead{Beam Matched \& Scaled} & \multirow{2}{*}{Equivalent ALMA Flux} & 
        \multirow{2}{*}{Percent Difference} & \multirow{2}{*}{Detection $\sigma$} \\
        & &CARMA Flux\tablenotemark{a}& & & \\
        & & (mJy beam$^{-1}$) & (mJy beam$^{-1}$) & &}
    \startdata
    \multirow{3}{*}{$uv$-plane Beam Matching}
        & 1  & 155.20 (25.47)  & 129.80 (0.85)  & -16.36 (13.74) & 1.19 \\
        & 10 & 966.61 (153.03) & 919.68 (5.59)  & -4.86 (15.07)  & 0.32 \\
        & 11 & 187.46 (34.39)  & 227.24 (5.59)  & 21.22 (22.44)  & 0.95 \\\hline\\
    \multirow{3}{*}{Simulated Re-Observation\tablenotemark{b}}
        & 1   & 134.89 (12.20) & 120.83 (0.01)  & -10.43 (8.10)  & 1.29 \\
        & 10  & 840.12 (66.76) & 860.11 (0.20)  & 2.38 (8.14)    & 0.29 \\
        & 11  & 162.93 (19.90) & 176.04 (0.20)  & 8.05 (13.19)   & 0.61   
    \enddata 
    \tablenotetext{a}{The flux for $uv$-plane Beam Matching and Simulated Re-observations differ 
        only in the rFCF from table \ref{tab:fcf_comp}.}
    \tablenotetext{b}{The uncertainty in the Equivalent ALMA flux measurements with simulated 
          re-observations is lower than in beam matching because the simulations do not include the 
          effects of noise; see section \ref{sec:alma_to_carma}.2}.   
    \label{tab:var_srcs_2}
\end{deluxetable*}

\begin{deluxetable*}{llccccccccc} %var_srcs_3
    \centering
    \tablewidth{0pt}
    \tablecolumns{10}
    \tablecaption{Variability of Sources, CARMA Track C2.3}
    \tablehead{\multirow{2}{*}{Beam Comparison Method}& \multirow{2}{*}{ID} &
        \colhead{Beam Matched \& Scaled} & \multirow{2}{*}{Equivalent ALMA Flux} & 
        \multirow{2}{*}{Percent Difference} & \multirow{2}{*}{Detection $\sigma$} \\
        & &CARMA Flux\tablenotemark{a}& & & \\
        & & (mJy beam$^{-1}$) & (mJy beam$^{-1}$) & &}
    \startdata
    \multirow{3}{*}{$uv$-plane Beam Matching}
        & 17   & 38.28 (14.75)   & 53.95 (0.23)   & 40.92 (54.29)   & 0.75 \\
        & 18   & 34.14 (13.89)   & 21.38 (0.24)   & -37.37 (25.50)  & 1.47 \\
        & 20   & 68.42 (21.34)   & 57.52 (0.22)   & -15.93 (26.22)  & 0.61 \\
        & 21   & 78.10 (26.51)   & 87.77 (0.30)   & 12.38 (38.14)   & 0.32 \\\hline\\  
    \multirow{3}{*}{Simulated Re-Observation\tablenotemark{b}}
        & 17   & 32.21 (12.95)   & 47.75 (0.11)   & 48.25 (59.59)   & 0.81 \\
        & 18   & 28.72 (12.14)   & 17.29 (0.11)   & -39.79 (25.46)  & 1.56 \\
        & 20   & 57.56 (19.13)   & 52.05 (0.01)   & -9.58 (30.05)   & 0.32 \\
        & 21   & 65.71 (23.54)   & 66.45 (0.15)   & 1.13 (36.23)    & 0.03
    \enddata 
    \tablenotetext{a}{The flux for $uv$-plane Beam Matching and Simulated Re-observations differ 
        only in the rFCF from table \ref{tab:fcf_comp}.}
        \tablenotetext{b}{The uncertainty in the Equivalent ALMA flux measurements with simulated 
            re-observations is lower than in beam matching because the simulations do not include 
            the effects of noise; see section \ref{sec:alma_to_carma}.2}. 
    \tablenotetext{a,b}{See table \ref{tab:var_srcs_1}.}   
    \label{tab:var_srcs_3}
\end{deluxetable*}

 %var_srcs tables, 3 of them

\section{Discussion \& Conclusion}
\label{sec:dis_and_conc}

Given that large bursts in deeply embedded protostars have only been detected a handful of times,
the lack of flux variations above the detection limits of $\sim24-180\%$ in our Serpens sample is
not unexpected. The only source in our sample with prior evidence of variability is Ser-emb 6
(SMM1), which is rising by $\sim5\%/yr$ in the Transient Survey \citep{johnstone2018}. Extrapolating
over the 9 year difference between the ALMA and CARMA epochs and assuming we would see a similar
change (ignoring differences in spatial scale and observing frequency) suggests we might have
expected a 45\% increase in flux, which would be at or above a 3$\sigma$ detection level for this 
source (see tables \ref{tab:var_srcs_1}-\ref{tab:var_srcs_3}). Given that we expect variability at 
the scales of the accretion disk, we would expect the signature of variability could be stronger 
than 45\% in the comparisons of ALMA and CARMA observations than in the Transient Survey where 
changes in flux are diluted by envelope material in the JCMT beam. Instead, the results of section 
\ref{sec:alma_to_carma}.4 are consistent with no change, suggesting that the rise in brightness of 
SMM1 seen by the Transient survey (between March 2016 and June 2017) may have only began recently.

The greatest limitations in our comparisons are imposed by the relatively low signal to noise of the
CARMA data and the small numbers of objects common to both the ALMA and CARMA observations available
for relative flux calibration, which hinder the determination of precise and statistically robust
rFCFs. Our comparisons of the ALMA data against itself however, suggest that if a second epoch with
similar resolution and sensitivity were obtained, variations at the level of a few percent could be
detected for sources with a SNR $> 100$, (those brighter than 10 mJy) about 14 in our sample.
Moreover, ALMA's excellent sensitivity makes such an observation efficient - our snapshot
observation reached a sensitivity of $100 \mu$ Jy in 40 minutes, compared to the sensitivity of the
CARMA maps of 1-3 mJy \citep{enoch2011}, achieved by combining data from over 20 nights of
observations over three years. As the JCMT Transient Survey finds 10\% of protostars varying at
$\sim\vert 5\vert \%$yr$^{-1}$, (including includes SMM1 (Ser-emb 6), an object also in our sample)
a second epoch of ALMA observations with a similar array configuration and integration time would
likely find robust low level variability in at least 1-2 objects. As the signature of variability is
likely being diluted by the JCMT beam at the scales of the envelope it probes, it is possible that
even more detections could be made.

The results of section \ref{sec:alma_to_carma} suggest $uv$-plane beam matching and simulated 
re-observations are similarly effective in terms of their sensitivity to variability, resulting 
rFCF precision, and ability to match the main lobes of the clean beams. While $uv$-plane beam 
matching is much simpler to implement, simulated re-observations can apply identical $uv$-plane 
sampling from one epoch of observations to the the other by carefully taking into account the array 
setup. Both methods could possibly be improved by taking into account differences in visibility 
weighting between two epochs in a more precise way than the $uv$-taper we have chosen to use. 
Future work should also focus on generalizing these methods to compare $>2$ epochs of observations, 
so reliable light curves can eventually be produced.

Some techniques not explored in this work could also improve the chances of finding variability. We 
have not taken into account small differences in observing frequency between our 233 GHz ALMA 
observations and 230 GHz CARMA observations. Assuming a spectral index for optically thin dust of 
2.5, a 3 GHz difference in frequency might see a difference in flux of 3.3\%. Our relative flux 
calibration should remove frequency dependent differences in dust emission in an average sense, but 
does not take into account variations in dust properties between objects. This can be corrected for 
by fitting a spectral index for each while performing deconvolution. This could be useful 
for comparing archival observations, which likely have been done with different observing 
frequencies as well as telescope configurations. Another unexplored technique is 
the possibility of measuring fluxes by fitting models of each source directly to the visibility 
data in the uv-plane. The structure of each source could be approximated using simple 
multi-component models, e.g. a point source embedded in a low-amplitude flat or Gaussian function 
to represent extended emission. This approach  
    would have the advantages of comparing the visibility data in a more direct way (i.e., without 
    deconvolution, which itself is essentially a model fitting process) and permitting all of the 
    data from different uv-plane samplings to be used instead of being culled or down-weighted, and 
    it could still be used in conjunction with a relative calibration scheme. Disadvantages of this 
    approach would include the modelling uncertainty in choice of functions used to represent each 
    source, and the necessity for a careful analysis of the visibility weights in order to ensure 
    robust uncertainties in the measured flux.
%Another unexplored technique is the possibility of 
%$uv$-plane model fitting methods for measuring fluxes. These have the advantage of comparing the 
%visibility data in the most direct way possible, but require a careful statistical analysis of how 
%the visibility weighting affects the uncertainty in the flux measurement.

Future work on identifying variability in deeply embedded protostars is underway. A Cycle 6 ALMA
proposal for 4 epochs of ACA-only Band 7 observations of variables in Serpens identified by the JCMT
Transient survey has been accepted (PI: Logan Francis, project code \texttt{2018.1.00917.S}). These 
observations will complement results from the contemporaneous Transient Survey by observing at 850 
$\mu$m with a resolution of 3.8\arcsec (compared to the 14.6\arcsec resolution of the JCMT), 
sufficient to reach the scale of the inner envelopes ($\sim 1500 $AU) of protostars in Serpens.

\section{Acknowledgements}

We would like to thank Helen Kirk, S{\"u}meyye Suri, Laura Perez, John Carpenter, and Gerald 
Scheiven for their support with ALMA and CARMA data reduction and useful insights on this project. 
We would also like to thank the anonymous referee for their helpful comments on 
this paper.

Doug Johnstone is supported by the National Research Council of Canada and by an NSERC Discovery
Grant. The National Radio Astronomy Observatory is a facility of the National Science Foundation
operated under agreement by the Associated Universities, Inc. ALMA is a partnership of ESO
(representing its member states), NSF (USA) and NINS (Japan), together with NRC (Canada) and NSC and
ASIAA (Taiwan) and KASI (Republic of Korea), in cooperation with the Republic of Chile. The Joint
ALMA Observatory is operated by ESO, AUI/ NRAO and NAOJ. This paper makes use of the following ALMA
data: ADS/JAO.ALMA\#2015.1.00310.S,

This work makes use of the following software: The Common Astronomy Software Applications (CASA) 
package \citep{casa2007}, CASA 
analysisUtils\footnote{https://safe.nrao.edu/wiki/bin/view/Main/CasaExtensions}, Python version 
2.7, astropy \citep{astropy2013}, aplpy \citep{aplpy2012}, 
and matplotlib \citep{matplotlib2007}.

\appendix

\section{Discussion of Individual Fields in Serpens Main} 

The Serpens Main cluster has been extensively surveyed at a variety of wavelengths and resolutions
over the past 30 years. Our ALMA observations provide some of the highest resolution and most
sensitive maps of deeply embedded protostars in the cluster to date. In many of our fields, we are
thus able to resolve single sources at the scale of the accretion disks, uncover significant
extended structure, and identify previously unknown faint sources. Here, we discuss each field in
the context of past and recent observations, and compare the positions of our sources with YSOs
identified in the Spitzer ``cores to disks'' (c2d) and ``Gould Belt'' (GB) surveys
\citep{dunham2015}. Table \ref{tab:dunham_sources} lists the properties of all c2d/GB YSOs in our 
fields associated with our mm sources in table \ref{tab:alma_sources}.

The Ser-emb objects targeted by our ALMA and the earlier CARMA observations \citep{enoch2011} were
originally defined from large scale Bolocam 1.1 mm and Spitzer mid-IR surveys, and classified
according to their bolometric temperature \citep{enoch2009}. Most of our targets lie in two dense
clusters: the northern Main Cluster (Ser-emb 4, 6, 8) and the southern Cluster B (Ser-emb 3, 7, 9,
11, 17). Three sources are relatively isolated (Ser-emb 2, 5, 15) from either cluster. An overview
of the region showing these clusters and the locations of the targeted embedded sources can be found
in figure 1 of \cite{enoch2011}. 

\begin{figure*}[h] %serpens_main_map
    \centering
    \includegraphics[scale=1.0]{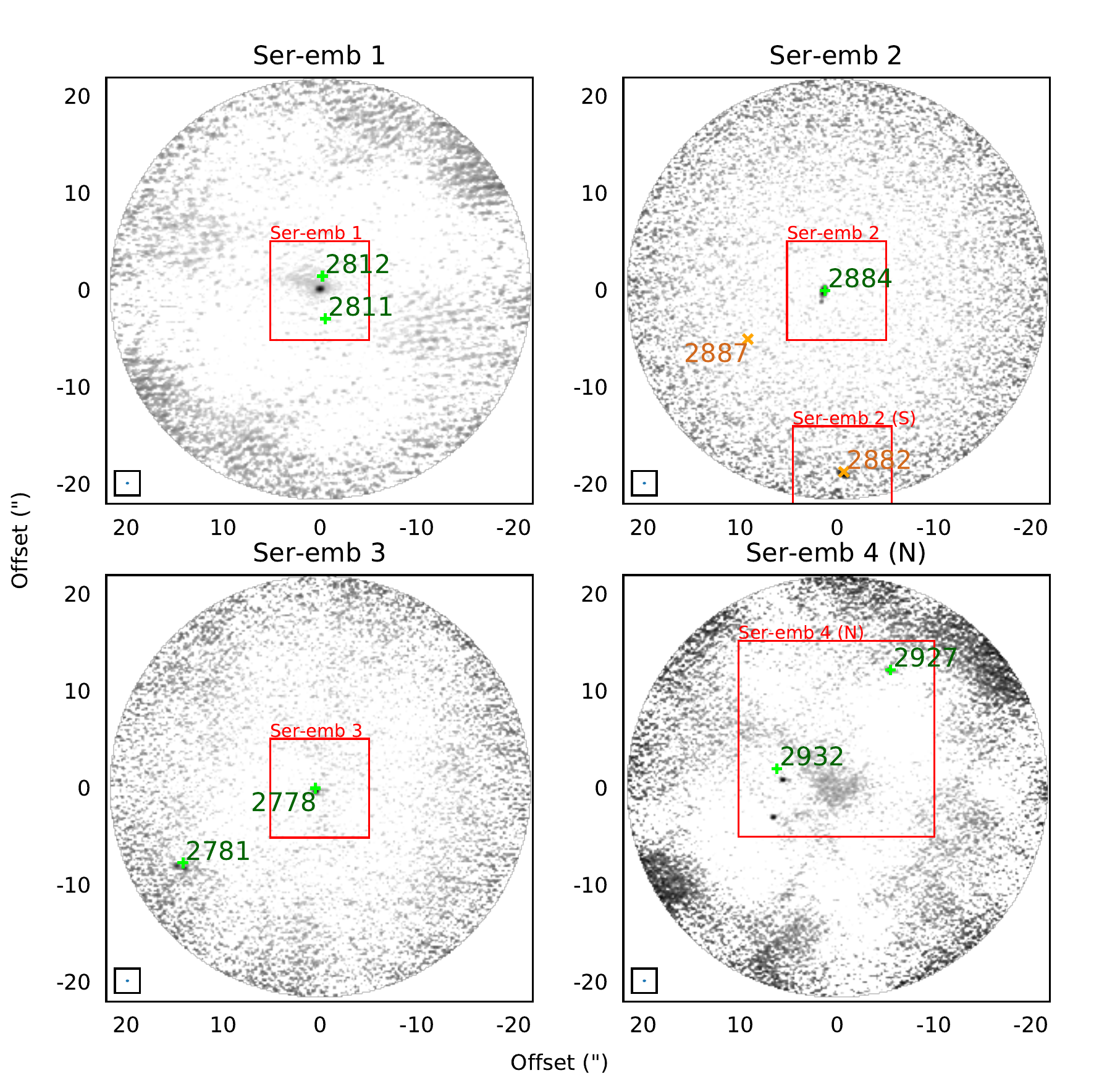}
    \caption{Full maps of our ALMA observations of Serpens protostars. Red squares indicate the 
        field of view for the postage stamps in figures \ref{fig:alma_serpens_maps_1} and 
        \ref{fig:alma_serpens_maps_2}. The maps are shown with primary beam correction to indicate 
        ALMA's field of view.  }  
    \label{fig:alma_serpens_finder_maps_1}
\end{figure*}

\begin{figure*}[htb] %alma_serpens_finder_maps_2
    \centering
    \includegraphics[scale=1.0]{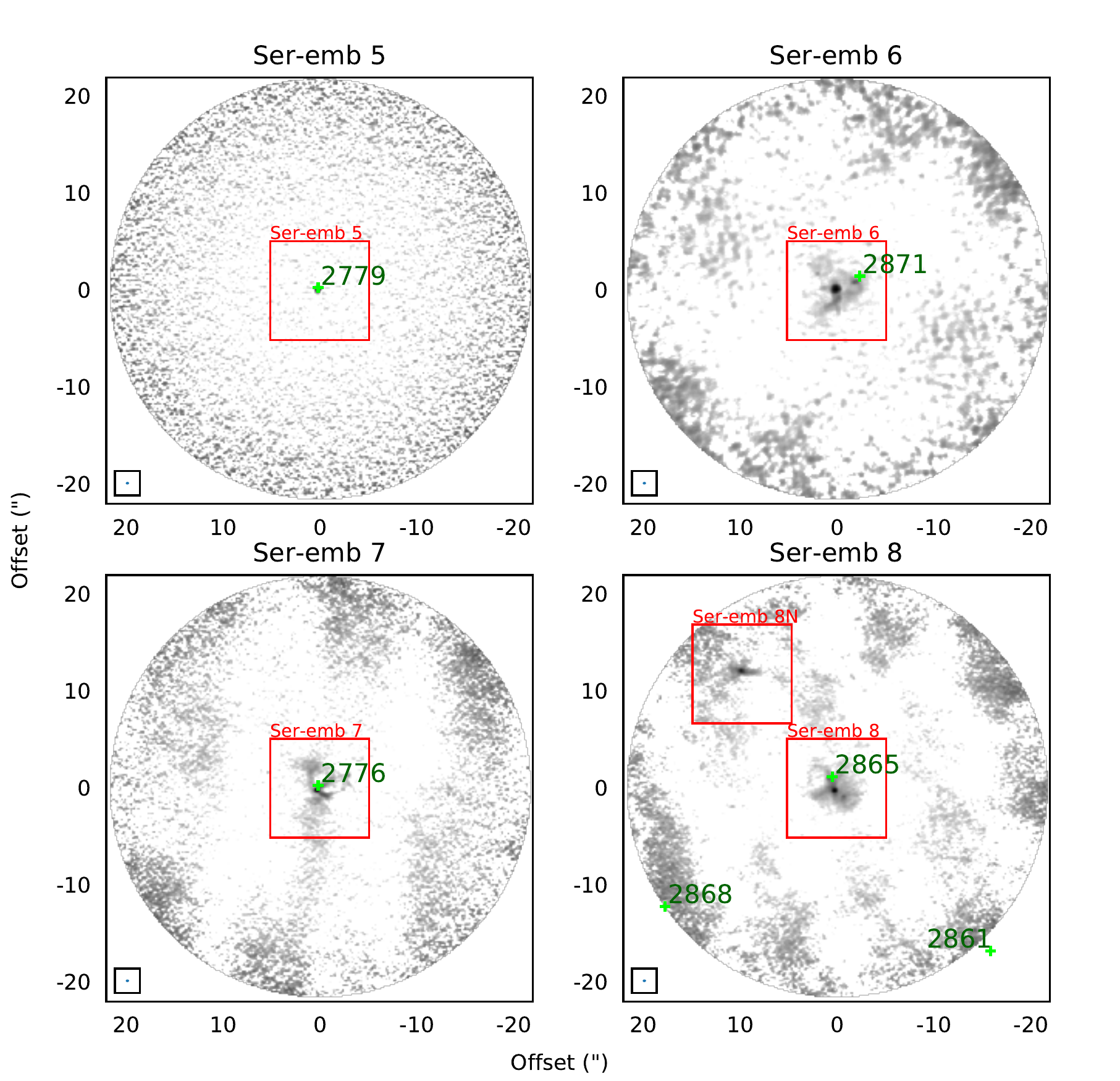}
    \caption{As figure \ref{fig:alma_serpens_finder_maps_1}.}  
    \label{fig:alma_serpens_finder_maps_2}
\end{figure*}

\begin{figure*}[htb] %alma_serpens_finder_maps_3
    \centering
    \includegraphics[scale=1.0]{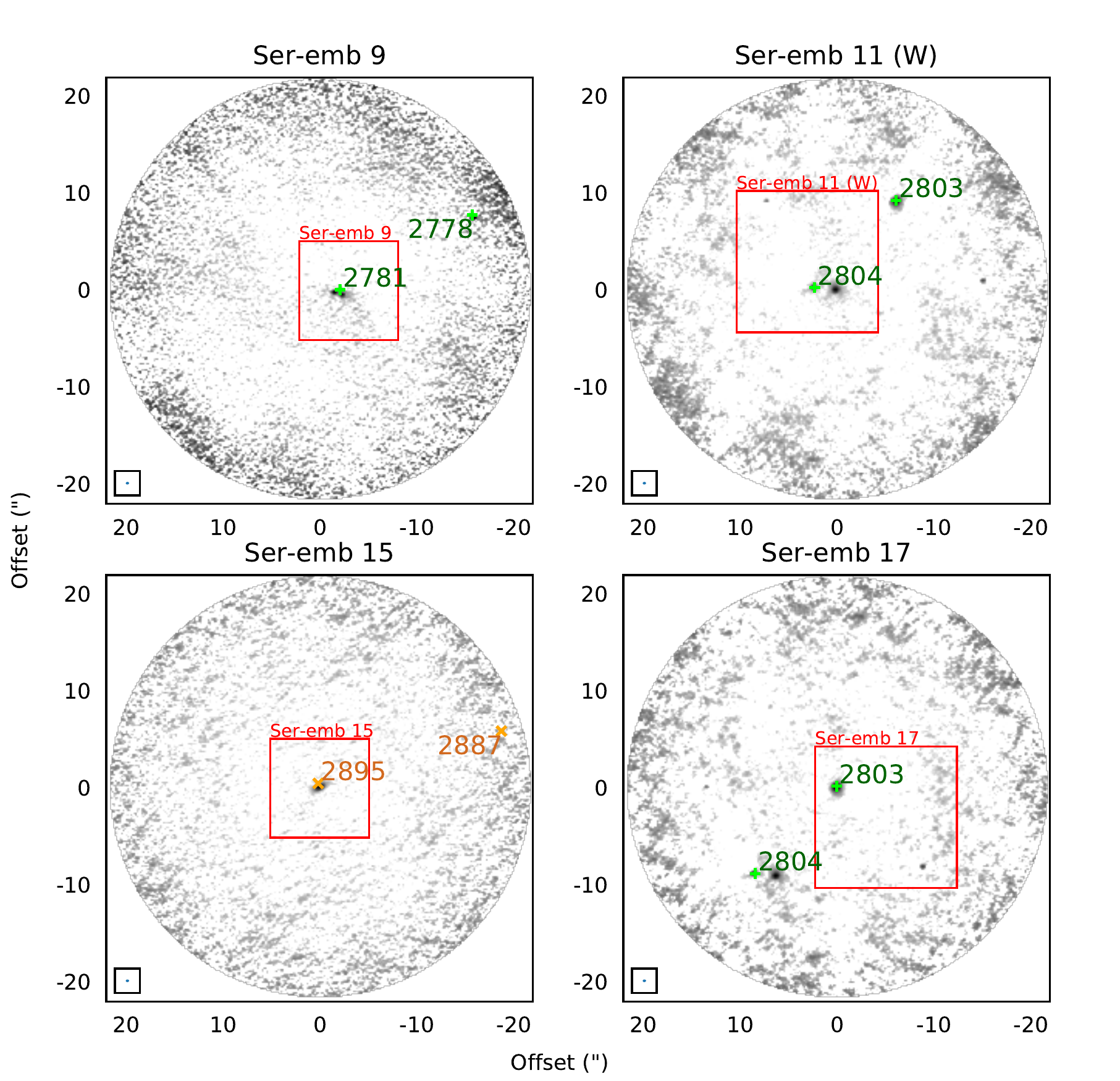}
    \caption{As figure \ref{fig:alma_serpens_finder_maps_1}.}  
    \label{fig:alma_serpens_finder_maps_3}
\end{figure*}

Our figures \ref{fig:alma_serpens_finder_maps_1}-\ref{fig:alma_serpens_finder_maps_3} show the 
continuum maps with the full field of view for each of our ALMA pointings, and indicate the 
positions and IDs of c2d/GB YSOs by green pluses (Class 0+1, Flat Spectrum) and orange crosses 
(Class II). Red squares show the location of the postage-stamp views in figures 
\ref{fig:alma_serpens_maps_1} and \ref{fig:alma_serpens_maps_2}.

\begin{comment}
\begin{itemize}
        \item Studies by Djupvik on multi-wavelength and Jet motion only look at the *South* 
        cluster, trace outflows to Ser-emb 1, 11, and 17. 
        \item JCMT Transient only looks at the North cluster.
        \item Chat Hull ALMA magnetic field studies only look at Ser-emb 6 and 8/8N.
        \item TADPOL has CARMA data in 1.3mm continuum and CO (J= 2-1) and SiO (J=5-4) lines (for 
        tracing outflows) for Ser-emb 1,6,8/8N.
\end{itemize}
\end{comment}

\begin{deluxetable*}{lccccccccc}
    \tabletypesize{\scriptsize}
    \tablewidth{0pt}
    \tablecaption{Properties of the Serpens c2d+GB YSOs}
    \tablehead{
        \colhead{} & \colhead{Spitzer} & \colhead{} & \multicolumn{3}{c}{Extinction Corrected} & \colhead{} & \colhead{Associated} & \colhead{}\\
        \colhead{} & \colhead{Source Name} & \colhead{$A_{\rm V}$} & \colhead{} & \colhead{\tbol$^{\prime}$} & \colhead{\lbol$^{\prime}$} & \colhead{} & \colhead{ALMA Source} & \colhead{Field} \\
        \colhead{Index} & \colhead{(SSTc2d or SSTgb +)} & \colhead{(mag)} & 
        \colhead{$\alpha^{\prime}$} & \colhead{(K)} & \colhead{(\lsun)} & 
        \colhead{Class\tablenotemark{a}} & \colhead{ID} & \colhead{Ser-emb \#} 
        }
    \startdata
    2776 & J182854.0+002930 &  9.6 &  1.15 &   69 &  8.80 & 0+I & 12     & Ser-emb 7         \\
    2778 & J182854.8+002952 &  9.6 &  1.60 &   51 &  6.60 & 0+I & 5      & Ser-emb 3,9       \\
    2779 & J182854.9+001832 &  9.6 &  0.68 &  150 &  0.14 & 0+I & 9      & Ser-emb 5         \\
    2781 & J182855.7+002944 &  9.6 &  1.65 &   26 &  1.60 & 0+I & 15,16  & Ser-emb 3,9       \\
    2803 & J182906.2+003043 &  9.6 &  1.33 &   67 & 10.00 & 0+I & 21     & Ser-emb 11(W), 17 \\
    2804 & J182906.7+003034 &  9.6 &  1.40 &   83 &  5.40 & 0+I & 17,18  & Ser-emb 11(W), 17 \\
    2811 & J182909.0+003128 &  9.6 &  0.07 &  420 &  0.03 & Flat& -      & Ser-emb 1         \\
    2812 & J182909.0+003132 &  9.6 &  2.13 &   36 &  4.20 & 0+I & 1      & Ser-emb 1         \\
    %2861 & J182947.0+011626 &  9.6 & -0.05 &  810 &  0.19 & Flat& -      & Ser-emb 8         \\
    2865 & J182948.1+011644 &  9.6 &  1.11 &   30 & 14.00 & 0+I & 14     & Ser-emb 8         \\
    %2868 & J182949.2+011631 &  9.6 &  1.38 &  410 &  0.95 & 0+I & -      & Ser-emb 8         \\
    2871 & J182949.6+011521 &  9.6 &  2.53 &   13 & 69.00 & 0+I & 11     & Ser-emb 6         \\
    2882 & J182952.3+003553 &  40.0& -1.55 & 2400 &  17.0 & II  & 4      & Ser-emb 2         \\
    2884 & J182952.5+003611 &  9.6 &  0.54 &   68 &  1.90 & 0+I & 2,3    & Ser-emb 2         \\
    2887 & J182953.0+003606 &  9.6 & -0.35 &  890 &  0.58 & II  & -      & Ser-emb 2, 15     \\
    2895 & J182954.3+003601 &  9.6 & -0.33 &   59 &  1.70 & II  & 20     & Ser-emb 15        \\
    2927 & J182959.9+011311 &  9.6 &  2.20 &  120 &  7.00 & 0+I & 6      & Ser-emb 4 (N)     \\
    2932 & J183000.7+011301 &  9.6 &  1.52 &   29 &  8.10 & 0+I & 7      & Ser-emb 4 (N)     
    \enddata
    \tablecomments{This table adapted from table 2 of \cite{dunham2015}.}    
    \tablenotetext{a}{Classes are defined by the
      extinction corrected spectral index $\alpha^\prime$ as Class 0+I: $\alpha'$\ $\geq
      0.3$, Flat-spectrum: $-0.3 \leq$ $\alpha'$\ $< 0.3$; Class II: $-1.6 \leq$
      $\alpha'$\ $< -0.3$; and Class III: $\alpha'$\ $< -1.6$ \citep{greene1994}.}    
\label{tab:dunham_sources}
\end{deluxetable*}
 %dunham_sources

\subsection{Ser-emb 1}

Ser-emb 1 is seen in our ALMA maps as a bright (100 mJy beam$^{-1}$) point-like source (ID 1) 
surrounded by faint, marginally detected ($3\sigma$) emission extending to the North-East. This
extended structure is likely a component of the bright emission visible on $\sim 10\arcsec$ scales
in the short-spacing CARMA maps of \cite{enoch2011}, which our ALMA observations are mostly
insensitive to.

Ser-emb 1 is the Class 0 source with the lowest bolometric temperature (39K) in Serpens Main
\citep{enoch2009}, suggesting it is also the least evolved. A N-S oriented bi-polar jet likely
originating from Ser-emb 1 is seen in 2.122 $\mu$m H$_2$ emission \citep{djupvik2016}. N-S oriented
CO outflows are also seen emanating directly from the source in CO (J = 2 $\rightarrow$ 1)
\cite{hull2014}.

Two c2d/GB YSOs \citep{dunham2015} are found within a few arcseconds of source 1. The closer
YSO, 2812, has a lower \tbol (36 K), similar to that found by \cite{enoch2009}. YSO 2811 is a warmer
(420 K) source with no corresponding mm emission visible in our ALMA maps, which is perhaps
unsurprising given its very low bolometric luminosity (0.03 \lsun) and more evolved flat spectrum
classification.

Ser-emb 1 is coincident (to within $\sim$ 2\arcsec) with a 3.6 cm radio continuum source detected 
by the VLA, which may result from thermal free-free emission in shocks \citep{djupvik2006}.

\subsection{Ser-emb 2}

Three compact, faint ($<10$mJy beam$^{-1}$) sources (IDs 2-4) are found in the maps of ALMA Ser-emb
2.  \cite{enoch2011} detected these sources only at a $5\sigma$ level in preliminary 110 GHz  maps,
but did not follow up with 230 GHz (1.3 mm) observations as was done for the other Ser-emb objects
due to their faintness.

The central sources in our map correspond to the location of Ser-emb 2 and consist of a disk-like
($\sim 275 $ AU major axis) component (ID 2) connected to a point source (ID 3) by a small ridge of
emission. The c2d/GB YSO 2884 is coincident with source 2 and has $\tbol=68$~K.

The source at the southern edge of the field (ID 4) also appears to be a compact resolved disk 
($\sim 257 $ AU major axis), and
is associated with c2d/GB YSO 2882. This class II YSO is more evolved and the hottest in our
sample, with a $\tbol=2400$~K. It is also the most optically extincted at $A_\text{V}=40$, compared
to the $A_\text{V}=9.6$ for all other YSOs in these fields.

Gould Belt YSO 2887 appears within the field, but with no corresponding mm emission, possibly 
because of its more evolved Class II status and lower luminosity ($\lbol=0.58L_\odot$). This YSO is 
also seen on the edge of the adjacent Ser-emb 15 field.

\subsection{Ser-emb 3 and 9}

Ser-emb 3 and 9 are located close enough together ($\sim15\arcsec$ apart) that they are both seen 
in two of our ALMA pointings. Neither were mapped at 230 GHz by \cite{enoch2011} due to the lack of 
clear detections in 110 GHz maps. 

Ser-emb 3 is seen as a faint $\sim9$ mJy beam$^{-1}$ point source (ID 5) associated with the c2d/GB 
YSO 2778, a $\tbol=51$~K class 0+I object.

Ser-emb 9 appears as a tight pair of $\sim9$ mJy beam$^{-1}$ peaks (IDs 15, 16) separated by just 
$\sim0.5\arcsec/215$~AU  and embedded in fainter emission. It is associated with the Gould Belt YSO 
2781, a Class 0+1 source with $\tbol=26$~K. Ser-emb 9 is also seen associated with a 3.6 cm radio 
to within 1\arcsec~\citep{djupvik2006}.

\subsection{Ser-emb 4 (N)}

ALMA observations of this field detect three faint ($<3$ mJy/beam) point sources (IDs 6, 7, 8). In
the CARMA observations, three regions of extended emission are detected and named Ser-emb 4 S, E,
and N. Ser-emb 4 N is the brightest component of this CARMA source, but due to its extended nature,
our ALMA configuration barely detects it at the field center. Our ALMA observations find the
Eastern-most point source (Source 8) is coincident with Ser-emb 4 E, but do not detect the fainter 
envelope around the source seen by CARMA. Ser-emb 4 S is similarly undetected.

The point source East of Ser-emb 4 N (Source 7) is associated with c2d/GB YSO 2932, a class 0+I 
source with $\tbol=28$~K. Since no compact mm emission or mid-IR c2d/GB sources are found at the 
positions of Ser-emb N and S, these sources are likely pre-stellar in nature.

The Northern-most point source (Source 6) in this field is identified as Ser-emb 19, a class I with 
$\tbol=129$~K in \cite{enoch2009}. It is also associated with Gould Belt 2927, a Class 0+I found to 
have a similar $\tbol=120$~K.

\subsection{Ser-emb 5}

 A single 7.8 mJy beam$^{-1}$ point source (ID 9) is found at the field center of our
ALMA observations. \cite{enoch2011} similarly detect a faint point source, and suggest the object is
the precursor to a brown dwarf or in a very low state of accretion due to its low luminosity
($\lbol=0.4\lsun$). The Class 0+I c2d/GB YSO 2779 is found at the position of this 
source, and has $\tbol=129$~K and a low luminosity ($\lbol=0.14L_\odot$. 

\subsection{Ser-emb 6}

Ser-emb 6 [also known as Serpens FIRS 1 \citep{harvey1984} and SMM 1 \citep{casali1993}] is the
brightest Class 0 source in Serpens Main and one of the most extensively studied. CARMA observations
of the source found an extended envelope surrounding two resolved sources. In our ALMA observations,
we see two extremely bright resolved sources ($\sim1000$~mJy~beam$^{-1}$, ID 10 and
$\sim100$~mJy~beam$^{-1}$, ID 11) surrounded by complex extended structure. For consistency with
other  high-resolution ALMA observations of this object [e.g \cite{hull2016}], we will refer to the
bright central source as SMM1-a and the relatively fainter western source as SMM1-b.

SMM1-a and b are associated with several jets and outflows. \cite{hull2016} find high velocity 
$\sim80$ km/s CO ($J=2\rightarrow1$) jets emanating from SMM1-a and b. They interpret the C-shaped 
extended structure around SMM1-a as walls of a cavity carved by precession of the jet. The same 
cavity is also seen in free-free emission from VLA observations, which \cite{hull2016} suggest to 
be caused by ionization of gas in shocks at the cavity walls. Polarization measurements with ALMA 
suggest that the jets are playing a role in shaping the local magnetic field \citep{hull2017a}. 
Lower velocity ($\sim10-20$ km/s) wide angle outflows are also seen in the CO ($J=2\rightarrow1$) 
emission around the high velocity jets \cite{hull2014,hull2017a}. Mid-IR Spitzer observations also 
find jets in H$_2$ and various atomic emission lines (e.g. [FeII]), however, interpreting which 
source is driving each outflow is complicated by the complexity of the outflows and lower Spitzer 
resolution \citep{dionatos2014}.

We find one c2d/GB YSO, 2871, coincident with SMM1-b, however, given that the beam size of
Spitzer ranges 2.5\arcsec to 40\arcsec depending on the instrument and wavelength, flux from the
brighter SMM1-a is almost certainly a large contribution if not dominating contribution to the YSO's
determined properties. Its low temperature ($\tbol=13$~K) and high luminosity ($\lbol=69L_\odot$)
agree well with the classification of Ser-emb 6 as a bright Class 0 source by \cite{enoch2009}. The 
coincidence of 2871 with source SMM1-a rather than b could also indicate that a is fainter at 
mid-IR wavelengths, but higher resolution observations would be needed to confirm this.

SMM1 is the only source in our sample which is confirmed to be variable at sub-mm wavelengths. It
has been rising in brightness by $\sim5 \%$yr$^{-1}$ in the first 18 months of the JCMT Transient 
Survey from December 2017 to June 2018 \citep{johnstone2018} and by $\sim2 \%$yr$^{-1}$ from 2012 
to 2016 in comparisons of archival Gould Belt Survey and Transient survey data \citep{mairs2017b}. 
Future epochs of ALMA observations should be able to determine if SMM1-a or b is the source of the 
rising brightness provided this trend continues.

\subsection{Ser-emb 7}

Ser-emb 7 is detected in the ALMA observations as a $\sim17$~mJy~beam$^{-1}$ point source (ID 12)
surrounded by complex and filamentary extended structure of $\sim1000$ AU in size. This is
suggestive of a fragmenting disk or interaction with outflows. No outflows in Spitzer maps of the
Cluster B region are linked to the structure surrounding Ser-emb 7, however, the source has yet to
be observed at high resolution in CO or another tracer. Ser-emb 7 and extended structure are also
seen in the CARMA observations, where maps constructed from large scale visibilities show an
envelope extending $\sim15\arcsec$ to the South of the source which is resolved out by our ALMA
configuration.

The Class 0+I c2d/GB YSO 2776 is associated with Ser-emb 7. It is found to have a 
($\tbol=60$K), similar to the ($\tbol=58$K) for Ser-emb 7 in \cite{enoch2009}. Ser-emb 7 is also 
associated with a 3.6~cm radio continuum source \cite{djupvik2006} 5\arcsec~to the North.

\subsection{Ser-emb 8/S68N}

Ser-emb 8 [Also known as S68N \citep{mcmullin1994} and SMM 9 \cite{casali1993}] is detected in our
ALMA maps as a $\sim30$~mJy~beam$^{-1}$ point source (ID 14) surrounded by knotty extended emission.
Another point source (ID 13) surrounded by extended structure is also detected to the North-East in 
our observations, here-after referred to as Ser-emb 8N. In the CARMA observations, large scale 
emission joins together the Ser-emb 8N and 8 in large scale maps. Both 8 and 8N power
molecular outflows observed in SiO ($J=5\rightarrow4$) extending SE-NE \citep{hull2014}. Maps of
this source in polarized dust emission find that magnetic fields at the 100-1000 AU scales are weak
and randomly oriented, suggesting turbulence plays a dominant role in establishing the field
morphology at these scales \cite{hull2017b}.

\cite{greene2018} have recently analysed a near-IR spectrum of Ser-emb 8 and detected features of 
the stellar photosphere (the first such detection and analysis for a Class 0 protostar), finding a 
photosphere temperature similar to pre-main-sequence stars, but with a lower surface gravity and 
larger stellar radius.

One class 0+1 Gould Belt YSO, 2865 is associated with Ser-emb 8, lying about 2\arcsec to the North 
of the bright central peak. No c2d/GB YSOs are associated with Ser-emb 8N, suggesting that it 
is too faint and/or deeply embedded to be detected at mid-IR wavelengths.  

\subsection{Ser-emb 11 (W) and 17}

Ser-emb 11 and 17 are located $\sim10\arcsec$ apart, and are thus seen in two of our ALMA pointings.
5 sources are found in both pointings (IDs 17-22). In both the ALMA and CARMA observations, Ser-emb 
11 and 17 are detected and Ser-emb 11 is resolved into two components (IDs 17, 18).  Both the 
targeted objects are bright, with peak fluxes of $\sim31$  and $\sim42$ ~mJy~beam$^{-1}$ for 
Ser-emb 11 (W) (ID 17 )and 17 (ID 21) respectively. Some extended emission is seen around Ser-emb 
11 and 17 in the ALMA maps.

Two previously unrecognized point sources (IDs 19, 22) are also found in the field. Both are faint,
with a peak flux of $\sim3$ (ID 19) and $\sim7$ (ID 22) ~mJy~beam$^{-1}$. Comparing the positions 
of these faint
sources with the large scale CARMA maps in figure 3 of \cite{enoch2011}, there is some extended
emission around the brighter $\sim7$~mJy~beam$^{-1}$ source directly West of Ser-emb 11 (W), but
none around the faint source North-East of Ser-emb 11 (W). The faintness of these sources makes them
qualitatively similar to Ser-emb 5, and thus they might also be proto-Brown dwarfs or objects in
very low level accretion states. However, no Gould Belt YSOs are associated with either object, nor
can we rule out the possibility we may be seeing a background sub-mm galaxy, making such an
interpretation insecure.

Ser-emb 11 and 17 have both been suggested as candidate driving sources for outflows seen in 2.122 
$\mu$m H$_2$ by Spitzer \cite{djupvik2016}. Outflows are also seen closer to each source in CO 
($J=2\rightarrow1$) \cite{hull2014}. 

Ser-emb 11 E (ID 18) is associated with c2d/GB YSO 2804, a $\tbol=67$K source, and Ser-emb 17 is  
similarly associated with YSO 2803, a $\tbol=73$K source. Similar bolometric temperatures are found 
by \cite{enoch2009}, who place both objects in Class I. Ser-emb 11 is additionally associated with 
a 3.6 cm continuum source less than $\sim 1$\arcsec~ away \cite{djupvik2006}.
  
\subsection{Ser-emb 15}

Ser-emb 15 is detected in both the ALMA and CARMA observations, and a disk-like ($\sim 126$ AU major
axis) source (ID 20) with a $\sim35$~mJy~beam$^{-1}$ peak is seen in our ALMA observations.

Ser-emb 15 is associated with c2d/GB YSO 2895, a marginal class II ($\alpha^\prime=-0.33$) 
source with  $\tbol=120$K, the warmest of our targeted Ser-emb objects. \cite{enoch2009} place 
Ser-emb 15 in Class I with $\tbol=100$K, which is likely a more appropriate categorization of the 
object given the extended disk seen in these ALMA observations.


\begin{thebibliography}{}
    \expandafter\ifx\csname natexlab\endcsname\relax\def\natexlab#1{#1}\fi
    
    \bibitem[{{Armitage} {et~al.}(2001){Armitage}, {Livio}, \&
        {Pringle}}]{armitage2001}
    {Armitage}, P.~J., {Livio}, M., \& {Pringle}, J.~E. 2001, \mnras, 324, 705
    
    \bibitem[{{Astropy Collaboration} {et~al.}(2013){Astropy Collaboration},
        {Robitaille}, {Tollerud}, {Greenfield}, {Droettboom}, {Bray}, {Aldcroft},
        {Davis}, {Ginsburg}, {Price-Whelan}, {Kerzendorf}, {Conley}, {Crighton},
        {Barbary}, {Muna}, {Ferguson}, {Grollier}, {Parikh}, {Nair}, {Unther},
        {Deil}, {Woillez}, {Conseil}, {Kramer}, {Turner}, {Singer}, {Fox}, {Weaver},
        {Zabalza}, {Edwards}, {Azalee Bostroem}, {Burke}, {Casey}, {Crawford},
        {Dencheva}, {Ely}, {Jenness}, {Labrie}, {Lim}, {Pierfederici}, {Pontzen},
        {Ptak}, {Refsdal}, {Servillat}, \& {Streicher}}]{astropy2013}
    {Astropy Collaboration}, {Robitaille}, T.~P., {Tollerud}, E.~J., {et~al.} 2013,
    \aap, 558, A33
    
    \bibitem[{{Audard} {et~al.}(2014){Audard}, {{\'A}brah{\'a}m}, {Dunham},
        {Green}, {Grosso}, {Hamaguchi}, {Kastner}, {K{\'o}sp{\'a}l}, {Lodato},
        {Romanova}, {Skinner}, {Vorobyov}, \& {Zhu}}]{audard2014}
    {Audard}, M., {{\'A}brah{\'a}m}, P., {Dunham}, M.~M., {et~al.} 2014, Protostars
    and Planets VI, 387
    
    \bibitem[{{Bae} {et~al.}(2014){Bae}, {Hartmann}, {Zhu}, \& {Nelson}}]{bae2014}
    {Bae}, J., {Hartmann}, L., {Zhu}, Z., \& {Nelson}, R.~P. 2014, \apj, 795, 61
    
    \bibitem[{{Bonnell} \& {Bastien}(1992)}]{bonnell1992}
    {Bonnell}, I., \& {Bastien}, P. 1992, \apjl, 401, L31
    
    \bibitem[{{Casali} {et~al.}(1993){Casali}, {Eiroa}, \& {Duncan}}]{casali1993}
    {Casali}, M.~M., {Eiroa}, C., \& {Duncan}, W.~D. 1993, \aap, 275, 195
    
    \bibitem[{{Cha} \& {Nayakshin}(2011)}]{cha2011}
    {Cha}, S.-H., \& {Nayakshin}, S. 2011, \mnras, 415, 3319
    
    \bibitem[{{Chen} {et~al.}(1995){Chen}, {Myers}, {Ladd}, \& {Wood}}]{chen1995}
    {Chen}, H., {Myers}, P.~C., {Ladd}, E.~F., \& {Wood}, D.~O.~S. 1995, \apj, 445,
    377
    
    \bibitem[{{Costigan} {et~al.}(2014){Costigan}, {Vink}, {Scholz}, {Ray}, \&
        {Testi}}]{costigan2014}
    {Costigan}, G., {Vink}, J.~S., {Scholz}, A., {Ray}, T., \& {Testi}, L. 2014,
    \mnras, 440, 3444
    
    \bibitem[{{Dempsey} {et~al.}(2013){Dempsey}, {Friberg}, {Jenness}, {Tilanus},
        {Thomas}, {Holland}, {Bintley}, {Berry}, {Chapin}, {Chrysostomou}, {Davis},
        {Gibb}, {Parsons}, \& {Robson}}]{dempsey2013}
    {Dempsey}, J.~T., {Friberg}, P., {Jenness}, T., {et~al.} 2013, \mnras, 430,
    2534
    
    \bibitem[{{Dionatos} {et~al.}(2014){Dionatos}, {J{\o}rgensen}, {Teixeira},
        {G{\"u}del}, \& {Bergin}}]{dionatos2014}
    {Dionatos}, O., {J{\o}rgensen}, J.~K., {Teixeira}, P.~S., {G{\"u}del}, M., \&
    {Bergin}, E. 2014, \aap, 563, A28
    
    \bibitem[{{Djupvik} {et~al.}(2006){Djupvik}, {Andr{\'e}}, {Bontemps}, {Motte},
        {Olofsson}, {G{\aa}lfalk}, \& {Flor{\'e}n}}]{djupvik2006}
    {Djupvik}, A.~A., {Andr{\'e}}, P., {Bontemps}, S., {et~al.} 2006, \aap, 458,
    789
    
    \bibitem[{{Djupvik} {et~al.}(2016){Djupvik}, {Liimets}, {Zinnecker}, {Barzdis},
        {Rastorgueva-Foi}, \& {Petersen}}]{djupvik2016}
    {Djupvik}, A.~A., {Liimets}, T., {Zinnecker}, H., {et~al.} 2016, \aap, 587, A75
    
    \bibitem[{{Dunham} {et~al.}(2010){Dunham}, {Evans}, {Terebey}, {Dullemond}, \&
        {Young}}]{dunham2010}
    {Dunham}, M.~M., {Evans}, II, N.~J., {Terebey}, S., {Dullemond}, C.~P., \&
    {Young}, C.~H. 2010, \apj, 710, 470
    
    \bibitem[{{Dunham} {et~al.}(2014){Dunham}, {Stutz}, {Allen}, {Evans},
        {Fischer}, {Megeath}, {Myers}, {Offner}, {Poteet}, {Tobin}, \&
        {Vorobyov}}]{dunham2014}
    {Dunham}, M.~M., {Stutz}, A.~M., {Allen}, L.~E., {et~al.} 2014, Protostars and
    Planets VI, 195
    
    \bibitem[{{Dunham} {et~al.}(2015){Dunham}, {Allen}, {Evans},
        {Broekhoven-Fiene}, {Cieza}, {Di Francesco}, {Gutermuth}, {Harvey},
        {Hatchell}, {Heiderman}, {Huard}, {Johnstone}, {Kirk}, {Matthews}, {Miller},
        {Peterson}, \& {Young}}]{dunham2015}
    {Dunham}, M.~M., {Allen}, L.~E., {Evans}, II, N.~J., {et~al.} 2015, \apjs, 220,
    11
    
    \bibitem[{{Enoch} {et~al.}(2009){Enoch}, {Evans}, {Sargent}, \&
        {Glenn}}]{enoch2009}
    {Enoch}, M.~L., {Evans}, II, N.~J., {Sargent}, A.~I., \& {Glenn}, J. 2009,
    \apj, 692, 973
    
    \bibitem[{{Enoch} {et~al.}(2011){Enoch}, {Corder}, {Duch{\^e}ne}, {Bock},
        {Bolatto}, {Culverhouse}, {Kwon}, {Lamb}, {Leitch}, {Marrone}, {Muchovej},
        {P{\'e}rez}, {Scott}, {Teuben}, {Wright}, \& {Zauderer}}]{enoch2011}
    {Enoch}, M.~L., {Corder}, S., {Duch{\^e}ne}, G., {et~al.} 2011, \apjs, 195, 21
    
    \bibitem[{{Greene} {et~al.}(2018){Greene}, {Gully-Santiago}, \&
        {Barsony}}]{greene2018}
    {Greene}, T.~P., {Gully-Santiago}, M.~A., \& {Barsony}, M. 2018, \apj, 862, 85
    
    \bibitem[{{Greene} {et~al.}(1994){Greene}, {Wilking}, {Andre}, {Young}, \&
        {Lada}}]{greene1994}
    {Greene}, T.~P., {Wilking}, B.~A., {Andre}, P., {Young}, E.~T., \& {Lada},
    C.~J. 1994, \apj, 434, 614
    
    \bibitem[{{Hartmann} {et~al.}(2016){Hartmann}, {Herczeg}, \&
        {Calvet}}]{hartmann2016}
    {Hartmann}, L., {Herczeg}, G., \& {Calvet}, N. 2016, \araa, 54, 135
    
    \bibitem[{{Hartmann} \& {Kenyon}(1996)}]{hartmann1996}
    {Hartmann}, L., \& {Kenyon}, S.~J. 1996, \araa, 34, 207
    
    \bibitem[{{Harvey} {et~al.}(1984){Harvey}, {Wilking}, \& {Joy}}]{harvey1984}
    {Harvey}, P.~M., {Wilking}, B.~A., \& {Joy}, M. 1984, \apj, 278, 156
    
    \bibitem[{{Herbig}(1977)}]{herbig1977}
    {Herbig}, G.~H. 1977, \apj, 217, 693
    
    \bibitem[{{Herbig}(2008)}]{herbig2008}
    ---. 2008, \aj, 135, 637
    
    \bibitem[{{Herczeg} {et~al.}(2017){Herczeg}, {Johnstone}, {Mairs}, {Hatchell},
        {Lee}, {Bower}, {Chen}, {Aikawa}, {Yoo}, {Kang}, {Kang}, {Chen}, {Williams},
        {Bae}, {Dunham}, {Vorobyov}, {Zhu}, {Rao}, {Kirk}, {Takahashi}, {Morata},
        {Lacaille}, {Lane}, {Pon}, {Scholz}, {Samal}, {Bell}, {Graves}, {Lee},
        {Parsons}, {He}, {Zhou}, {Kim}, {Chapman}, {Drabek-Maunder}, {Chung},
        {Eyres}, {Forbrich}, {Hillenbrand}, {Inutsuka}, {Kim}, {Kim}, {Kuan}, {Kwon},
        {Lai}, {Lalchand}, {Lee}, {Lee}, {Long}, {Lyo}, {Qian}, {Scicluna}, {Soam},
        {Stamatellos}, {Takakuwa}, {Tang}, {Wang}, \& {Wang}}]{herczeg2017}
    {Herczeg}, G.~J., {Johnstone}, D., {Mairs}, S., {et~al.} 2017, \apj, 849, 43
    
    \bibitem[{{Hodapp}(1999)}]{hodapp1999}
    {Hodapp}, K.~W. 1999, \aj, 118, 1338
    
    \bibitem[{{Hodapp} {et~al.}(2012){Hodapp}, {Chini}, {Watermann}, \&
        {Lemke}}]{hodapp2012}
    {Hodapp}, K.~W., {Chini}, R., {Watermann}, R., \& {Lemke}, R. 2012, \apj, 744,
    56
    
    \bibitem[{{Holland} {et~al.}(2013){Holland}, {Bintley}, {Chapin},
        {Chrysostomou}, {Davis}, {Dempsey}, {Duncan}, {Fich}, {Friberg}, {Halpern},
        {Irwin}, {Jenness}, {Kelly}, {MacIntosh}, {Robson}, {Scott}, {Ade},
        {Atad-Ettedgui}, {Berry}, {Craig}, {Gao}, {Gibb}, {Hilton}, {Hollister},
        {Kycia}, {Lunney}, {McGregor}, {Montgomery}, {Parkes}, {Tilanus}, {Ullom},
        {Walther}, {Walton}, {Woodcraft}, {Amiri}, {Atkinson}, {Burger}, {Chuter},
        {Coulson}, {Doriese}, {Dunare}, {Economou}, {Niemack}, {Parsons},
        {Reintsema}, {Sibthorpe}, {Smail}, {Sudiwala}, \& {Thomas}}]{holland2013}
    {Holland}, W.~S., {Bintley}, D., {Chapin}, E.~L., {et~al.} 2013, \mnras, 430,
    2513
    
    \bibitem[{{Hull} {et~al.}(2014){Hull}, {Plambeck}, {Kwon}, {Bower},
        {Carpenter}, {Crutcher}, {Fiege}, {Franzmann}, {Hakobian}, {Heiles}, {Houde},
        {Hughes}, {Lamb}, {Looney}, {Marrone}, {Matthews}, {Pillai}, {Pound},
        {Rahman}, {Sandell}, {Stephens}, {Tobin}, {Vaillancourt}, {Volgenau}, \&
        {Wright}}]{hull2014}
    {Hull}, C.~L.~H., {Plambeck}, R.~L., {Kwon}, W., {et~al.} 2014, \apjs, 213, 13
    
    \bibitem[{{Hull} {et~al.}(2016){Hull}, {Girart}, {Kristensen}, {Dunham},
        {Rodr{\'{\i}}guez-Kamenetzky}, {Carrasco-Gonz{\'a}lez}, {Cort{\'e}s}, {Li},
        \& {Plambeck}}]{hull2016}
    {Hull}, C.~L.~H., {Girart}, J.~M., {Kristensen}, L.~E., {et~al.} 2016, \apjl,
    823, L27
    
    \bibitem[{{Hull} {et~al.}(2017{\natexlab{a}}){Hull}, {Girart}, {Tychoniec},
        {Rao}, {Cort{\'e}s}, {Pokhrel}, {Zhang}, {Houde}, {Dunham}, {Kristensen},
        {Lai}, {Li}, \& {Plambeck}}]{hull2017a}
    {Hull}, C.~L.~H., {Girart}, J.~M., {Tychoniec}, {\L}., {et~al.}
    2017{\natexlab{a}}, \apj, 847, 92
    
    \bibitem[{{Hull} {et~al.}(2017{\natexlab{b}}){Hull}, {Mocz}, {Burkhart},
        {Goodman}, {Girart}, {Cort{\'e}s}, {Hernquist}, {Springel}, {Li}, \&
        {Lai}}]{hull2017b}
    {Hull}, C.~L.~H., {Mocz}, P., {Burkhart}, B., {et~al.} 2017{\natexlab{b}},
    \apjl, 842, L9
    
    \bibitem[{{Hunter}(2007)}]{matplotlib2007}
    {Hunter}, J.~D. 2007, Computing in Science and Engineering, 9, 90
    
    \bibitem[{{Hunter} {et~al.}(2006){Hunter}, {Brogan}, {Megeath}, {Menten},
        {Beuther}, \& {Thorwirth}}]{hunter2006}
    {Hunter}, T.~R., {Brogan}, C.~L., {Megeath}, S.~T., {et~al.} 2006, \apj, 649,
    888
    
    \bibitem[{{Hunter} {et~al.}(2017){Hunter}, {Brogan}, {MacLeod}, {Cyganowski},
        {Chandler}, {Chibueze}, {Friesen}, {Indebetouw}, {Thesner}, \&
        {Young}}]{hunter2017}
    {Hunter}, T.~R., {Brogan}, C.~L., {MacLeod}, G., {et~al.} 2017, \apjl, 837, L29
    
    \bibitem[{{Jensen} \& {Haugb{\o}lle}(2018)}]{jensenhaugbolle2018}
    {Jensen}, S.~S., \& {Haugb{\o}lle}, T. 2018, \mnras, 474, 1176
    
    \bibitem[{{Johnstone} {et~al.}(2013){Johnstone}, {Hendricks}, {Herczeg}, \&
        {Bruderer}}]{johnstone2013}
    {Johnstone}, D., {Hendricks}, B., {Herczeg}, G.~J., \& {Bruderer}, S. 2013,
    \apj, 765, 133
    
    \bibitem[{{Johnstone} {et~al.}(2018){Johnstone}, {Herczeg}, {Mairs},
        {Hatchell}, {Bower}, {Kirk}, {Lane}, {Bell}, {Graves}, {Aikawa}, {Chen},
        {Chen}, {Kang}, {Kang}, {Lee}, {Morata}, {Pon}, {Scicluna}, {Scholz},
        {Takahashi}, {Yoo}, \& {The JCMT Transient Team}}]{johnstone2018}
    {Johnstone}, D., {Herczeg}, G.~J., {Mairs}, S., {et~al.} 2018, \apj, 854, 31
    
    \bibitem[{{Kenyon} {et~al.}(1990){Kenyon}, {Hartmann}, {Strom}, \&
        {Strom}}]{kenyon1990}
    {Kenyon}, S.~J., {Hartmann}, L.~W., {Strom}, K.~M., \& {Strom}, S.~E. 1990,
    \aj, 99, 869
    
    \bibitem[{{Liu} {et~al.}(2017){Liu}, {Dunham}, {Pascucci}, {Bourke}, {Hirano},
        {Longmore}, {Andrews}, {Carrasco-Gonz{\'a}lez}, {Forbrich},
        {Galv{\'a}n-Madrid}, {Girart}, {Green}, {Ju{\'a}rez}, {K{\'o}sp{\'a}l},
        {Manara}, {Palau}, {Takami}, {Testi}, \& {Vorobyov}}]{liu2017}
    {Liu}, H.~B., {Dunham}, M.~M., {Pascucci}, I., {et~al.} 2017, ArXiv e-prints,
    arXiv:1710.08686
    
    \bibitem[{{Lodato} \& {Clarke}(2004)}]{lodato2004}
    {Lodato}, G., \& {Clarke}, C.~J. 2004, \mnras, 353, 841
    
    \bibitem[{{Machida} {et~al.}(2011){Machida}, {Inutsuka}, \&
        {Matsumoto}}]{machida2011}
    {Machida}, M.~N., {Inutsuka}, S.-i., \& {Matsumoto}, T. 2011, \apj, 729, 42
    
    \bibitem[{{Mairs} {et~al.}(2017{\natexlab{a}}){Mairs}, {Lane}, {Johnstone},
        {Kirk}, {Lacaille}, {Bower}, {Bell}, {Graves}, {Chapman}, \& {The JCMT
            Transient Team}}]{mairs2017a}
    {Mairs}, S., {Lane}, J., {Johnstone}, D., {et~al.} 2017{\natexlab{a}}, \apj,
    843, 55
    
    \bibitem[{{Mairs} {et~al.}(2017{\natexlab{b}}){Mairs}, {Johnstone}, {Kirk},
        {Lane}, {Bell}, {Graves}, {Herczeg}, {Scicluna}, {Bower}, {Chen}, {Hatchell},
        {Aikawa}, {Chen}, {Kang}, {Kang}, {Lee}, {Morata}, {Pon}, {Scholz},
        {Takahashi}, {Yoo}, \& {The JCMT Transient Team}}]{mairs2017b}
    {Mairs}, S., {Johnstone}, D., {Kirk}, H., {et~al.} 2017{\natexlab{b}}, \apj,
    849, 107
    
    \bibitem[{{McKee} \& {Offner}(2011)}]{mckeeoffner2011}
    {McKee}, C.~F., \& {Offner}, S.~R.~R. 2011, in IAU Symposium, Vol. 270,
    Computational Star Formation, ed. J.~{Alves}, B.~G. {Elmegreen}, J.~M.
    {Girart}, \& V.~{Trimble}, 73--80
    
    \bibitem[{{McMullin} {et~al.}(1994){McMullin}, {Mundy}, {Wilking}, {Hezel}, \&
        {Blake}}]{mcmullin1994}
    {McMullin}, J.~P., {Mundy}, L.~G., {Wilking}, B.~A., {Hezel}, T., \& {Blake},
    G.~A. 1994, \apj, 424, 222
    
    \bibitem[{{McMullin} {et~al.}(2007{\natexlab{a}}){McMullin}, {Waters},
        {Schiebel}, {Young}, \& {Golap}}]{mcmullin2007}
    {McMullin}, J.~P., {Waters}, B., {Schiebel}, D., {Young}, W., \& {Golap}, K.
    2007{\natexlab{a}}, in Astronomical Society of the Pacific Conference Series,
    Vol. 376, Astronomical Data Analysis Software and Systems XVI, ed. R.~A.
    {Shaw}, F.~{Hill}, \& D.~J. {Bell}, 127
    
    \bibitem[{{McMullin} {et~al.}(2007{\natexlab{b}}){McMullin}, {Waters},
        {Schiebel}, {Young}, \& {Golap}}]{casa2007}
    {McMullin}, J.~P., {Waters}, B., {Schiebel}, D., {Young}, W., \& {Golap}, K.
    2007{\natexlab{b}}, in Astronomical Society of the Pacific Conference Series,
    Vol. 376, Astronomical Data Analysis Software and Systems XVI, ed. R.~A.
    {Shaw}, F.~{Hill}, \& D.~J. {Bell}, 127
    
    \bibitem[{{Myers} \& {Ladd}(1993)}]{myers1993}
    {Myers}, P.~C., \& {Ladd}, E.~F. 1993, \apj, 413, L47
    
    \bibitem[{{Nayakshin} \& {Lodato}(2012)}]{Nayakshin2012}
    {Nayakshin}, S., \& {Lodato}, G. 2012, \mnras, 426, 70
    
    \bibitem[{{Ortiz-Le{\'o}n} {et~al.}(2017){Ortiz-Le{\'o}n}, {Dzib}, {Kounkel},
        {Loinard}, {Mioduszewski}, {Rodr{\'{\i}}guez}, {Torres}, {Pech}, {Rivera},
        {Hartmann}, {Boden}, {Evans}, {Brice{\~n}o}, {Tobin}, \&
        {Galli}}]{ortiz-leon2017}
    {Ortiz-Le{\'o}n}, G.~N., {Dzib}, S.~A., {Kounkel}, M.~A., {et~al.} 2017, \apj,
    834, 143
    
    \bibitem[{{Pfalzner} {et~al.}(2008){Pfalzner}, {Tackenberg}, \&
        {Steinhausen}}]{pfalzner2008}
    {Pfalzner}, S., {Tackenberg}, J., \& {Steinhausen}, M. 2008, \aap, 487, L45
    
    \bibitem[{{Plunkett} {et~al.}(2015){Plunkett}, {Arce}, {Mardones}, {van
            Dokkum}, {Dunham}, {Fern{\'a}ndez-L{\'o}pez}, {Gallardo}, \&
        {Corder}}]{plunkett2015}
    {Plunkett}, A.~L., {Arce}, H.~G., {Mardones}, D., {et~al.} 2015, \nat, 527, 70
    
    \bibitem[{{Rab} {et~al.}(2017){Rab}, {Elbakyan}, {Vorobyov}, {G{\"u}del},
        {Dionatos}, {Audard}, {Kamp}, {Thi}, {Woitke}, \& {Postel}}]{rab2017}
    {Rab}, C., {Elbakyan}, V., {Vorobyov}, E., {et~al.} 2017, \aap, 604, A15
    
    \bibitem[{{Reipurth}(1990)}]{reipurth1990}
    {Reipurth}, B. 1990, in IAU Symposium, Vol. 137, Flare Stars in Star Clusters,
    Associations and the Solar Vicinity, ed. L.~V. {Mirzoian}, B.~R. {Pettersen},
    \& M.~K. {Tsvetkov}, 229--251
    
    \bibitem[{{Remijan} {et~al.}(2015){Remijan}, {Adams}, {Eiji}, \&
        {Andreani}}]{almatechcycle3}
    {Remijan}, A., {Adams}, M., {Eiji}, A., \& {Andreani}, P. 2015, ALMA Technical
    Handbook Cycle 3
    
    \bibitem[{{Robitaille} \& {Bressert}(2012)}]{aplpy2012}
    {Robitaille}, T., \& {Bressert}, E. 2012, {APLpy: Astronomical Plotting Library
        in Python}, Astrophysics Source Code Library, ascl:1208.017
    
    \bibitem[{{Safron} {et~al.}(2015){Safron}, {Fischer}, {Megeath}, {Furlan},
        {Stutz}, {Stanke}, {Billot}, {Rebull}, {Tobin}, {Ali}, {Allen}, {Booker},
        {Watson}, \& {Wilson}}]{safron2015}
    {Safron}, E.~J., {Fischer}, W.~J., {Megeath}, S.~T., {et~al.} 2015, \apjl, 800,
    L5
    
    \bibitem[{{Sault} {et~al.}(1995){Sault}, {Teuben}, \& {Wright}}]{sault1995}
    {Sault}, R.~J., {Teuben}, P.~J., \& {Wright}, M.~C.~H. 1995, in Astronomical
    Society of the Pacific Conference Series, Vol.~77, Astronomical Data Analysis
    Software and Systems IV, ed. R.~A. {Shaw}, H.~E. {Payne}, \& J.~J.~E.
    {Hayes}, 433
    
    \bibitem[{{Shu} {et~al.}(1987){Shu}, {Adams}, \& {Lizano}}]{shu1987}
    {Shu}, F.~H., {Adams}, F.~C., \& {Lizano}, S. 1987, \araa, 25, 23
    
    \bibitem[{{Simon} {et~al.}(2011){Simon}, {Hawley}, \& {Beckwith}}]{simon2011}
    {Simon}, J.~B., {Hawley}, J.~F., \& {Beckwith}, K. 2011, \apj, 730, 94
    
    \bibitem[{{Taquet} {et~al.}(2016){Taquet}, {Wirstr{\"o}m}, \&
        {Charnley}}]{taquet2016}
    {Taquet}, V., {Wirstr{\"o}m}, E.~S., \& {Charnley}, S.~B. 2016, \apj, 821, 46
    
    \bibitem[{{Tassis} \& {Mouschovias}(2005)}]{tassis2005}
    {Tassis}, K., \& {Mouschovias}, T.~C. 2005, \apj, 618, 783
    
    \bibitem[{{Venuti} {et~al.}(2015){Venuti}, {Bouvier}, {Irwin}, {Stauffer},
        {Hillenbrand}, {Rebull}, {Cody}, {Alencar}, {Micela}, {Flaccomio}, \&
        {Peres}}]{venuti2015}
    {Venuti}, L., {Bouvier}, J., {Irwin}, J., {et~al.} 2015, \aap, 581, A66
    
    \bibitem[{{Vorobyov} \& {Basu}(2005)}]{vorobyov2005}
    {Vorobyov}, E.~I., \& {Basu}, S. 2005, \apjl, 633, L137
    
    \bibitem[{{Vorobyov} \& {Basu}(2006)}]{vorobyov2006}
    ---. 2006, \apj, 650, 956
    
    \bibitem[{{Vorobyov} \& {Basu}(2010)}]{vorobyov2010}
    ---. 2010, \apj, 719, 1896
    
    \bibitem[{{Yoo} {et~al.}(2017){Yoo}, {Lee}, {Mairs}, {Johnstone}, {Herczeg},
        {Kang}, {Kang}, {Cho}, \& {The JCMT Transient Team}}]{yoo2017}
    {Yoo}, H., {Lee}, J.-E., {Mairs}, S., {et~al.} 2017, \apj, 849, 69
    
    \bibitem[{{Zauderer} {et~al.}(2016){Zauderer}, {Bolatto}, {Vogel}, {Carpenter},
        {Per{\'e}z}, {Lamb}, {Woody}, {Bock}, {Carlstrom}, {Culverhouse}, {Curley},
        {Leitch}, {Plambeck}, {Pound}, {Marrone}, {Muchovej}, {Mundy}, {Teng},
        {Teuben}, {Volgenau}, {Wright}, \& {Wu}}]{zauderer2016}
    {Zauderer}, B.~A., {Bolatto}, A.~D., {Vogel}, S.~N., {et~al.} 2016, \aj, 151,
    18
    
    \bibitem[{{Zhu} {et~al.}(2009{\natexlab{a}}){Zhu}, {Hartmann}, \&
        {Gammie}}]{zhu2009a}
    {Zhu}, Z., {Hartmann}, L., \& {Gammie}, C. 2009{\natexlab{a}}, \apj, 694, 1045
    
    \bibitem[{{Zhu} {et~al.}(2010){Zhu}, {Hartmann}, \& {Gammie}}]{zhu2010}
    ---. 2010, \apj, 713, 1143
    
    \bibitem[{{Zhu} {et~al.}(2009{\natexlab{b}}){Zhu}, {Hartmann}, {Gammie}, \&
        {McKinney}}]{zhu2009b}
    {Zhu}, Z., {Hartmann}, L., {Gammie}, C., \& {McKinney}, J.~C.
    2009{\natexlab{b}}, \apj, 701, 620
    
\end{thebibliography}
\end{document}